\documentclass[twocolumn,runinaddress,frontmatterverbose,aps, showkeys, showpacs,prb,floatfix]{revtex4-1}
\usepackage{graphicx}
\usepackage{subfigure}
\usepackage{dcolumn}
\usepackage{bm}
\usepackage{graphicx}
\usepackage{color}
\usepackage{colortbl}
\usepackage[dvipsnames]{xcolor} 
\usepackage{tabularx}
\usepackage{epsfig}
\usepackage{amsmath}
\usepackage{amssymb}
\usepackage{graphicx}
\usepackage{dcolumn,epstopdf}
\usepackage{bm}
\usepackage{ctable}
\definecolor{grn}{rgb}{0,0,0.54}

\usepackage{hyperref} 
\hypersetup{
    colorlinks=true,       
    linkcolor=blue,        
    citecolor=blue,        
    filecolor=blue,     
    urlcolor=blue    
}
\usepackage{latexsym}
\usepackage{amssymb}
\usepackage{amsfonts}
\usepackage{amsmath}
\def\be{\begin{equation}}
\def\ee{\end{equation}}
\newcommand{\bea}{\begin{eqnarray}}
\newcommand{\eea}{\end{eqnarray}}

\newcommand{\ang}{\text{\normalfont\AA}}

\begin{document}


\title{DFT Calculations of Temperature-Dependent NQR Parameters in $\alpha$-paradichlorobenzene and $\beta$-HMX}

\author{Allen R. Majewski}
\email{majewski@phys.ufl.edu}
\author{Chris R. Billman}
\author{Hai-Ping Cheng}
\author{N. S. Sullivan}
\affiliation{ Department of Physics, University of Florida, Florida 32611, USA. \\ 
}

%
%
%
%

\begin{abstract}
A method for first principles predictions of observed temperature-dependent NQR spectra is presented using density functional theory (DFT) and the isobaric $T$-dependent NQR frequencies of $^{35}$Cl and $^{14}$N nuclei are computed for the two molecular crystals (1) $\alpha$-paradichlorobenzene, and (2) the nitroamine high explosive $\beta$-HMX ($\beta$-octahydro-1,3,5,7-tetranitro-1,3,5,7-tetrazocine) over a range of 200 K and up to room temperature. Notably, the method requires no supposition of a form for the intermolecular potentials or wave functions, requires no particular insight as to the nature of the internal motions or of the chemical bonds present, and does not depend on the crystal structure, making the method amenable to any periodic solid for which experimentally determined structural data are available. For each substance, unit cells of various volume are prepared using experimentally determined atomic positions and cell parameters. In each of the prepared volume-corrected cells, a molecular dynamics (MD) simulation generates a set of perturbed atomic positions, the collection of which is intended to represent the system at a given $T,V$. For each configuration of atoms generated along the MD trajectory, the electric field gradient (EFG) tensors are computed at the site of each quadrupolar nucleus. The rotational displacements of the moving EFG principal axes from their equilibrium directions are used to apply a dynamic correction to the DFT-computed static-lattice NQR frequencies, resulting in first-principles DFT predictions of $T$-dependent NQR spectra at constant pressure. Because the $V$-dependence from thermal expansion and the $T$-dependence due to internal motions are simultaneously considered, the NQR's notoriously model-dependent temperature coefficients are computed entirely ab-initio.
\end{abstract}
\date{\today}
\maketitle

%
%
%
%
%
%
%
%
%
%
%
\section{Introduction}

Nuclear quadrupole resonance (NQR) spectroscopy is a radio-frequency (RF) technique for solid materials in which a quadrupolar nucleus is irradiated with RF magnetic fields and the resulting transitions between the nuclear spin energy levels are directly observed. \cite{pound1950nuclear}  NQR arises from the interaction between a nuclear electric quadrupole moment and the local electric field gradient (EFG) that is natural to the system.  Because the EFG, a second-rank tensor, depends on the particular charge distribution in the space surrounding the nucleus, NQR is a sensitive probe of the electronic structure of systems.  As such, NQR experiments have long been used to carry out basic investigations on the structure and dynamics of a wide range of solid materials. \cite{dodgen1956observation} \cite{kind1976cl} \cite{karpowicz1983librational} \cite{karpowicz1984comparison}  

Calculations of NQR parameters can assist in the interpretation of NQR spectra and help overcome practical difficulties of carrying out NQR measurements.  The inability to predict NQR unknown frequencies presents the greatest challenge for NQR experimentation in new materials.  This study presents a method to overcome this challenge by allowing direct calculations of room temperature NQR frequencies that are sufficiently accurate to narrow the search space for NQR exploration in new materials.

\subsection{NQR Applications and Limitations}
Because NQR frequencies in a material result from the substance's unique electronic structure, detection of the material-specific NQR spectra can unambiguously indicate the presence of a particular substance \cite{das1958nuclear} and can be used to identify or differentiate materials.  Exploitation of ths "chemical fingerprint" property of NQR spectra has long been considered in the development of technology of considerable practical use.  NQR is known to be capable of identifying counterfeit pharmaceuticals \cite{dass2011anti}.  Additionally, NQR has a long history of consideration for remote and non-invasive detection of materials such as landmines, bombs, \cite{buess1993explosives} \cite{smith1995nitrogen} \cite{mozjoukhine2000two} \cite{gregorovivc2009tnt} and narcotics \cite{garroway1994narcotics} \cite{magnuson2001system}, as is evidenced by the patent record of the past three decades. The United States government has extensively characterized  NQR spectra of energetic materials like TNT and HMX. \cite{marino1982nitrogen} Yet, despite the obvious merits of a point-and-shoot bomb detector, and the substantial body of NQR research conducted to this end, no such device has been shown to work in practice. Nevertheless, development of NQR continues to this day in the pursuit of miniaturization and automation of detector operation, including "smart" devices running system-on-chip operating systems and other technologies \cite{ariando2019autonomous}.  

While there is continuous development of NQR technology for its potential commercial use, NQR is much less common than NMR \cite{vij2007handbook} studies in basic solid state research due to practical difficulties of the technique. For the inductive pickups used in NQR and NMR detection, signal-to-noise ratio (SNR) increases with frequency due to Faraday's law of induction. Yet the naturally arising NQR frequencies are often quite low: most NQR frequencies fall below 100MHz but can be as low as several kHz \cite{lucken1969nuclear} in the case of $^{14}N$ due to its small quadrupole moment. Consequently, NQR studies can have prohibitively poor (SNR) \cite{das1958nuclear}.  NMR studies often have much higher SNR because the Larmor frequency $\omega_{L} = \gamma B_{ext}$, and hence the SNR, can be made very high by increasing the applied magnetic field $B_{ext}$, where $\gamma$ is the gyromagnetic ratio of nucleus.  

Additionally, NQR is often shelved in favor of NMR because typical NQR detectors are easily reused as NMR detectors.  Because they are determined by the system's natural EFG, which is a consequence of the system's particular electronic structure, NQR frequencies are material specific;  observed NQR frequencies cannot be chosen or appreciably modified by an experimenter.  As such, the operating frequency of an NQR detector is determined by the sample itself. The low SNR in NQR studies necessitates the receiver section of an NQR detector to have very high sensitivity in a narrow passband to accommodate typical NQR line-widths of a few kHz.  Due to the impedance matching needs of the sample coil pickup circuitry, the consequences of lossy substrates, and other challenges in the construction of necessary low noise amplifiers and probe circuits NQR detectors that are capable of broadband operation is a challenging engineering task.  As a consequence, NQR experiments are often carried out using home-built "one-off" spectrometers, useful for only a sample under consideration, or at best, some other similar materials.  Development of off-the-shelf, reusable NQR spectrometers remains an active area of research for commercial use, however. \cite{dass2011anti}

The principal reason NQR methods are not used as extensively as NMR for investigating the properties of materials is the inability to predict the frequencies accurately.  Even approximate theoretical prediction of NQR frequencies would make NQR easier to carry out.  This is because for new materials, in which NQR has never been performed, locating unknown NQR requires a search over a range of plausible NQR frequencies as they are not known \textit{a-priori} to the experimenter.  Low SNR and ignorance of the NQR frequencies make NQR impractical or impossible if the search space is not sufficiently constrained, as one may not observe the NQR frequencies at all. Accurate calculations of unknown NQR frequencies by computer simulation would mitigate the problem of locating unknown NQR lines, reducing the barrier to entry for NQR use to probe interesting physics in new materials.

Recent developments in DFT codes allow calculation of NQR frequencies that are valid for $T=0$ K. \cite{bonhomme2010new} However NQR experiments (or any experiment, for that matter) are more easily performed at room temperature, making reliable prediction of room temperature NQR frequencies highly desirable.  A method for making such predictions with a straight-forward, broadly applicable procedure for use in arbitrary materials is presented here.

\subsection{GIPAW NQR Predictions and Limitations}

Introduced in 2001 to compute NMR chemical shifts, the gauge-including projector augmented-wave (GIPAW) method \cite{pickard2001all} allows calculations of electric field gradients and corresponding static-lattice NQR frequencies in the plane wave pseudopotential formalism of density-functional theory (DFT) \cite{yates2007computations} \cite{bonhomme2010new} \cite{charpentier2011paw} \cite{bonhomme2012first}.  However, the strong temperature dependence of NQR frequencies, which arises primarily due to internal motions, can not be ascertained using the vanilla DFT routines that are currently available because the GIPAW method, like many routines based on DFT, can only be applied to a static-lattice undergoing no internal motions.  Without consideration of lattice dynamics, the results of GIPAW calculated NQR frequencies are only valid for $T=0$ K.  Yet, all the practical commercial uses of NQR detection in materials detection and characterization would seem to require room temperature operation; after all, airports tend to be kept at room temperature and pharmaceutical use cases would be more realistic without the need for cryogenics.  Even for basic science needs, NQR as a probe of interesting transitions across a temperature range involves slowly cooling down the sample from higher temperatures.  Because NQR frequencies can vary hundreds of kHz or more across a temperature range of 300 K, static lattice calculations do not sufficiently narrow the search space for finding unknown NQR frequencies.  

 GIPAW NQR calculations that are performed in the static lattice necessarily neglect lattice dynamics, so should not be expected to match room temperature NQR frequencies. Furthermore, unless the atomic positions and lattice vectors used in the GIPAW calculation were taken at the same temperature as the NQR experiment, the often significant volume dependence of the NQR frequencies is also neglected by the calculations. 
     
Presently, no GIPAW study known to this author accounts for the strong temperature dependence of NQR frequencies.  Previous work on the development of GIPAW calculations of NQR frequencies is often performed in the static-lattice in a variety of structures. \cite{socha2017exploring} \cite{milinkovic2012nmr} \cite{peric2014solid}  \cite{leroy2019121} The calculations not only neglect the effect of motions, but furthermore, if the simulation's cell volume does not match that of the NQR experiment's, then the calculations are performed on a system that is perhaps very different from the experimental one.  The relevance of these calculations to room temperature NQR data is unclear, though the comparison is often made. Thermal expansion can have a number of effects on NQR frequencies.  If the structure is not optimized with the correct lattice constants, a DFT simulation would not faithfully represent the ``experimental static-lattice NQR frequency`` - that is, the NQR one \textit{would} measure in a hypothetical frozen lattice that has the cell volume of the room temperature sample. Not even the fitting parameter meant by ``static-lattice NQR frequency``, which is never directly observed, is computed as advertised if care is not taken to prepare the cell to faithfully represent the structure studied in the experiment.

\subsection{Inclusion of Motional Effects}
Consideration of lattice dynamics in conjunction with DFT calculations is indeed a fundamental problem.  The effect of internal motions on DFT calculated NMR shielding tensors has been investigated using vibrational averaging in both solid materials \cite{lee2007high} \cite{rossano2005first} as well as gases and solutions  \cite{tang2007vibrational}  \cite{bagno2007computing}  \cite{waller200851v}  \cite{rohrig2008nmr} \cite{kongsted2007nuclear}  \cite{straka2008toward}  \cite{harding2008quantitative}.  Molecular dynamics (MD) simulations have been used in conjunction with GIPAW to include the influence of internal motions in DFT calculated NMR chemical shifts in organic solid materials \cite{dumez2009calculation}, but no method has been proposed to compute temperature dependent NQR parameters in materials from first principles, but no method has been proposed to account for the effect of internal motions on observed NQR using first principles DFT calculations. In this study, the plane-wave DFT code QUANTUM ESPRESSO \cite{QE-2009} \cite{QE-2017} is used to determine the isobaric $T$-dependence of $^{35}$Cl and $^{14}$N NQR frequencies in $\alpha$-pCl$_{2}\phi$ and $\beta$-HMX, respectively.  

By accounting for the influence of internal motions and the volume dependence of the electronic structure simultaneously, the T-dependent NQR spectra and corresponding BBK model temperature coefficients are predicted at atmospheric pressure in both materials using the same procedure.  Notably, this method does not require any consideration of the vibrational spectra, nor of the nature of the structure and chemical bonding in the system. Taking as inputs only the relevant crystallographic structural information of atomic positions and the effect of thermal expansion on lattice vectors, the method is both structure-agnostic and model-independent making its application to arbitrary systems straight-forward.

%
%
%
%
%
%
%
%
%
%
%
%
\section{Theory}
%
%
%
%
\subsection{NQR Transition Frequencies in the Static Lattice}
Nuclei with a non-spherical nuclear charge density $\rho(\mathbf{x})$ can have an electric quadrupole moment \\${Q=\frac{1}{e} \int ( 3 z^2 - r^2)\rho(\mathbf{x})d^3 x}$ that interacts with the external electric potential $\phi$ due to its electronic environment via the electric field gradient (EFG)

\begin{equation}
(\nabla \mathbf{E})^{ij} = \frac{\partial^2 \phi}{\partial x_i \partial x_j} = \phi_{ij}
\end{equation}
Similar to the manner of NMR, in which the interaction of the nuclear magnetic dipole moment with an external magnetic field can be probed by irradiation of the nucleus with RF of a particular frequency, NQR frequencies result from the splitting of the nuclear spin energy levels caused by this interaction.  NQR frequencies depend on the quadrupole coupling constant $C_{q}$ and the asymmetry parameter $\eta$ defined as 

\begin{equation}
C_{q} = \frac{e Q\phi_{ZZ}}{h}
\end{equation}

\begin{equation}
\eta = \frac{\phi_{YY} - \phi_{XX}}{\phi_{ZZ}}
\end{equation}
where  $\phi_{ii}$ are the eigenvalues of the EFG tensor at the nuclear site.  The directions of the eigenvectors define the principle axes of the EFG tensor, the names of which are chosen by the convention $|\phi_{ZZ}| \ge |\phi_{YY}| \ge |\phi_{XX}|$ ensuring $0 \le \eta \le 1$.  NQR transition frequencies are then expressed in terms of only $C_{q}$ and $\eta$, though the form of the expression will depend on the nuclear spin of the particular isotope.  For spin $I=\frac12$, $Q=0$ and there are no transition frequencies. If $I > \frac12$, the NQR frequencies are proportional to $C_{q}$ and a factor in $\eta$ which can have a significant impact on $\nu_{q}$ when $\eta$ approaches unity.

$^{35}$Cl has nuclear spin $I=\frac32$ in which case there is just one NQR frequency

\begin{equation}\label{nqrspin32}
\nu = \frac12 C_{q} \sqrt{1+ \frac{\eta^2}{3}}
\end{equation}
In the $I=1$ case of $^{14}$N, there can be one, two, or three transitions for $I=1$ depending on the magnitude of $\eta$. The frequencies are given by 

\begin{equation}\label{nqrspin1}
\nu_{\pm} = \frac34 C_{q} \left( 1 \pm \frac{\eta}{3} \right)
\end{equation}

\begin{equation}\label{delta_nu}
\nu_{\delta} = \nu_{+} - \nu_{-} = \frac{\eta C_{q}}{2}
\end{equation}
Note that as $\eta$ approaches zero, $\nu_{+} \to \nu_{-}$ and $\nu_{\delta} \to 0$.  As $\eta$ approaches unity, $\nu_{\delta} \to \nu_{-} = \frac12 \nu_{+}$.

%
%

%
%
\subsection{Motional Effects and T-dependence}\label{sec:motional_volume}

As a nucleus is displaced due to motions within a crystal, it is subjected to a rapidly changing electronic environment. The nuclear position changes at a high frequency within the spatially dependent EFG.  Furthermore, EFG is also time dependent in the constantly evolving system undergoing motions. The EFG seen by the nucleus therefore has a somewhat intractable dependence on space and time.  Yet observed NQR frequencies are not time dependent because the motional frequencies in solids are many orders of magnitude higher than the NQR transition frequencies.  The effective observed $C_{q}$ in a system with a dynamic lattice is determined by the time averaged EFG to which the nucleus is subjected in a system undergoing motion. The nature of the internal motions change significantly with temperature as the magnitude of the nuclear displacements and number of modes increase with with increasing $T$. The internal motions that lead to the time averaged effective EFG at the nuclear site are the cause of NQR's strict $T$-dependence.  The time averaged EFG can be shown to have a weaker associated $C_{q}$ than that of the equilibrium structure, meaning internal motions act to decrease the observed NQR frequency. 

While the effect of motions is normally to reduce the NQR frequencies, the nature of the $V$-dependence due to thermal expansion (an implicit $T$-dependence) is more subtle. If the expansion is highly anisotropic, as is the case in many monoclinic crystals, both $C_{q}$ and $\eta$ are affected by changes in the magnitudes and directions of the lattice vectors.  Thermal expansion can also cause reorientation of molecules within the cells of molecular crystals, which may affect the NQR parameters in any fashion, either gradually or abruptly. Moreover, in molecular crystals, intermolecular distances increase during expansion of the cell, while intramolecular atomic distances do not appreciably change. As a result, increasing molecular separation in crystal cells gives rise to EFG's with larger associated magnitudes of $C_{q}$ as the electronic environment of the nucleus possesses greater anisotropy in a lone molecule than in a crystal cell with neighboring molecules.  This means that in molecular crystals, $C_{q}$ and the NQR frequency $\nu_{q}$ at constant $T$ strictly increase with increasing $V$. On the contrary,  ionic crystals display the opposite trend because the limiting case as $V \to \infty$ is for the nucleus to be alone in space, making the local EFG vanish. Because of the complexity of NQR $V$-dependence, previous models for the precise nature of the isobaric NQR $T$-dependence vary between structures as the volume correction to observed EFGs can in general have any affect on $\nu_{q}$.  In this work, a method that is generally applicable to materials irrespective of their nature is proposed to account for all these effects.  By performing a particular sequence of DFT calculations resulting in the calculation of the correct $C_{q}$ and $\eta$, along with a dynamic normalization factor $\kappa_T$ which accounts for internal motions, the computed parameters are sensitive to both $T$ and $V$, allowing one to predict laboratory NQR $T$-dependence.

\subsection{Pioneering Work}
A general observation from early NQR research is that the rate of change of the NQR frequency with respect to temperature $\frac{d\nu}{dT}$ tends to zero at low temperature and becomes a negative constant as $T$ increases to 293 K and beyond. \cite{brown1990anomalous}.  It was found that the character of the $T$-dependence differed in ionic versus molecular crystals.  Furthermore, the values of $C_q$ in a molecular crystal were typically around 5\% lower than the ones observed in single molecules.  Attribution of these effects to underlying physical processes, disentangling their dependence on $T$ and $V$, accounting for otherwise anomalous behavior of $T$-dependent NQR frequencies became the subject of many theoretical works in the following decades which carefully consider the dynamics and the thermodynamic properties of individual systems.

The Bayer model is a theoretical framework for the $T$-dependence of NQR frequencies that was first reported by Horst Bayer in 1951 \cite{bayer1951theorie}.  The concepts in the paper underpin many similar models which are currently in use that are essentially extensions to his work accounting for additional effects.  Bayer proposed that the internal motions of the system are responsible for the temperature curves observed in NQR experiments by considering the rotations of the EFG principle axes in a system undergoing motion.  Paradichlorobenzene was used to illustrate this concept in his original paper and it is shown that internal motions result in a reduced NQR frequency compared with that which would be observed in a static lattice.  Bayer arrived at a functional form for NQR frequencies which captures the desired limiting behavior 

\begin{equation}\label{bayer-fit}
\nu_{q}\left(T\right) = a - \frac{b}{\exp\left(c/T\right) - 1}
\end{equation}
where $a$, $b$, and $c$ are parameters to be fit.  The model considers a single model of librations for the paradichlorobenze molecule. Bayer's theory had some success but because it is a constant volume theory, it failed to account for the variation of $\nu_{q}$ across larger temperature ranges if thermal expansion significantly impacts the static lattice EFG.  
 
Highly influential experimental and theoretical work by Kushida, Benedeck, Bloombergen \cite{kushida1956dependence} reported in 1956 presented KBB model which extends the ideas of Bayer to include arbitrary numbers of vibrational modes and the effect thermal expansion on the parameter $C_q$ \footnote{In the KBB paper as well as many works, the parameter $C_q$ is referred to instead by $q = \phi_{ZZ}/e$} and the resulting NQR frequency.  The KBB paper emphasizes the thermodynamic considerations affecting NQR frequencies by separately considering the dependence of NQR parameters temperature, pressure, and volume. The KBB paper reported very thorough measurements of the pressure dependence of NQR frequencies in three polymorphs of paradichlorobenzene at several temperatures.  Their work elucidated the complexity of the general problem of modeling NQR's isobaric $T$-dependence by disentangling the various competing effects of each $T,V,P$, as well as distinguishing ionic and molecular crystals by including measurements $P$-dependence NQR frequencies at constant T in potassium chlorate (KClO$_{3}$) and cuprous oxide (Cu$_{2}$O).  The KBB model proposes a functional form 

\begin{equation}\label{KBB-fit}
\nu_{q}\left(T\right) = \nu_{0} \left(1 + b^{\prime} T + \frac{c^{\prime}}{T}\right) = a + b T + \frac{c}{T}
\end{equation}
where $\nu_0 = a$ refers to the NQR frequency computed by equations \eqref{nqrspin32} and \eqref{nqrspin1} using the values of $C_q$ and $\eta$ at $T=0$.  KBB acknowledges the volume dependence of $C_{q}$ and $\eta$, which the authors addressed by assuming an equation of state for the solid in order to construct $\nu$-$vs$-$V$ plots.  Their work emphasized the relationship of NQR parameters to thermodynamic conditions, characterizing the individual dependence on $P,V$ and $T$ of observed NQR frequencies in several materials. 

The effect on $\nu_{q}$ due to the $V$-dependence of the amplitudes of lattice vibrations was carried out \cite{brown1960temperature} shortly after giving rise to the Brown model in which a quadratic term in $T^2$ is added to the KBB fitting function. The seminal works of Bayer, KBB, and Brown have stood the test of time as their early considerations underpin many existing variants of their original functional forms for $\nu_{q}(T)$, which primarily offer corrections for additional effects and special cases.  A summary of just some of the many available fitting functions is presented in one of the early attempts to study NQR frequencies using DFT by simulating the vibrational spectra of molecules in order to directly calculate the fitting parameters for the Bayer, KBB, and Brown models, as well some of their many descendants. \cite{latosinska2002studies}

\subsection{The Dynamic Normalization Factor $\kappa_{T}$}
In this work, the many effects considered separately by the Bayer, KBB, and Brown models are captured at once by determination of a parameter $\kappa_T$ which accounts for internal motions, and by explicit calculation the $V$-dependence of $C_{q}$ and $\eta$.  Calculation of the isobaric $T$-dependence of NQR frequencies amounts to calculation of $\kappa_T$, $C_{q}(V)$, and $\eta(V)$.  It is worthwhile to review the original assumptions of Bayer which underpin the theoretical basis of this work. Clear descriptions of these essential considerations were given by Hahn \cite{das1958nuclear} and Lucken shortly later \cite{lucken1969nuclear}. 

Consider a nucleus that is displaced from its equilibrium position in a system undergoing motion.  The displaced nucleus will experience at its new location an EFG tensor with principle axes $X$, $Y$, and $Z$ having orientations that are rotated with respect to the directions of the equilibrium principle axes $X^{\prime}$, $Y^{\prime}$, $Z^{\prime}$.

 The rotated tensor components can be expressed in the space-fixed coordinate frame as \cite{das1958nuclear} \cite{lucken1969nuclear}

\begin{eqnarray}\label{tensors_thetas}
\phi_{X^\prime X^\prime} &=  &\left( 1 - \theta_{Y}^2 - \theta_{Z}^2 \right)\phi_{XX} +   \theta_{Z}^2 \phi_{YY}  + \theta_{Y}^2 \phi_{ZZ}          \nonumber  \\
\phi_{Y^\prime Y^\prime} &=  &\theta_{X}^2 \phi_{XX}   + \left( 1 - \theta_{Y}^2 - \theta_{Z}^2 \right)\phi_{YY}  + \theta_{X}^2 \phi_{ZZ}          \nonumber  \\
\phi_{Z^\prime Z^\prime} &=  &\theta_{Y}^2 \phi_{XX}  + \theta_{Z}^2 \phi_{YY}   + \left( 1 - \theta_{Y}^2 - \theta_{Z}^2 \right)\phi_{ZZ}                              \\
\phi_{X^\prime Y^\prime} &= &\theta_{Z}\phi_{XX} + \left( \theta_{X}\theta_{Y} - \theta_{Z} \right) \phi_{YY} - \theta_{X}\theta_{Y}\phi_{ZZ}    \nonumber  \\
\phi_{Y^\prime Z^\prime} &= &-\theta_{Y}\theta_{Z}\phi_{XX} + \theta_{X} \phi_{YY} + \left( \theta_{Y}\theta_{Z} - \theta_{X} \right) \phi_{ZZ}  \nonumber \\
\phi_{Z^\prime X^\prime} &= &-\theta_{Y}\phi_{XX} - \theta_{X}\theta_{Z} \phi_{YY} + \left( \theta_{X}\theta_{Z} + \theta_{Y} \right) \phi_{ZZ}  \nonumber 
\end{eqnarray}
where $\theta_i$ are intrinsic rotations about the $i^{th}$ axis through which the stationary equilibrium principle axes system could be mapped to the moving axes.  It is proposed that the observed coupling constant for a system at non-zero temperature will be given by an average of the $\phi_{i^{\prime}j^{\prime}}$ in equation \eqref{tensors_thetas} over the time-dependent angular displacements $\theta_{X}, \theta_{Y}$ and $\theta_{Z}$ \cite{lucken1969nuclear} the system develops as a result of its various modes internal motion. Presuming the motions are a superposition of simple oscillators, the average angular displacements vanish,

\begin{equation} \langle \theta_{i} \rangle = 0 \end{equation}
averages of the squared angular displacements will survive in equation \eqref{tensors_thetas}. Therefore the averaged tensor is diagonal and it can be shown that the $T$-dependent coupling constant is given by the static lattice value of the coupling constant $C_{q,0}$ and a dynamic normalization factor $\kappa_{T}$, that is depends on both $V$ and $T$, which accounts for the effect internal motions 
\begin{equation}\label{kappa_times_Cq}
C_{q}\left(T\right) = \kappa_{T} C_{q,0}
\end{equation} 
where $C_{q,0}$ is the volume ($V$) dependent static-lattice value of the coupling constant$C_{q}$ and $\kappa_{T}$ is given by \cite{wang1955pure}

\begin{equation} \label{kappaT}
	\kappa_{T} = \Big( 1 - \frac32 \left(\langle \theta_{X}^2\rangle + \langle \theta_{Y}^2\rangle \right) - \frac12 \eta  \left(\langle \theta_{X}^2\rangle - \langle \theta_{Y}^2\rangle \right) \Big) 
\end{equation}
The NQR frequencies are always proportional to $C_{q}$ and therefore are adjusted by the same factor $\kappa_{T}$,

\begin{equation}\label{kappaproduct}
\nu \left( T \right) = \kappa_{T}  \nu_{0}
\end{equation}
where $\nu_{0}$ is the NQR frequency calculated using the static lattice formulas \eqref{nqrspin32} or \eqref{nqrspin1} and $C_{q,0}$. 

$C_{q,0}$ and $\eta$ will, of course, generally depend on volume in a complex manner due to overall changes in the structure and orientation of the system and it?s constituent nuclei or molecules, was noted in section \ref{sec:motional_volume}.  The subtlety of volume dependence and its net effect on NQR frequencies is the primary roadblock in to developing a model for isobaric $T$-dependence that applies to general materials and temperature ranges.  Equations \eqref{kappa_times_Cq} and \eqref{kappaproduct} imply that the isobaric $T$-dependent NQR frequencies can be computed exactly if $C_{q,0}$, $\eta$, and $\kappa_{T}$, and their dependence on $T,V$ are known.

Because $\langle \theta_{i}^{2}\rangle > 0$, and $\kappa_{T}$ is reduced by $\frac32 \left(\langle \theta_{X}^2\rangle + \langle \theta_{Y}^2\rangle \right)$, which dominates the term in $\eta$ and $\langle \theta_{X}^2\rangle - \langle \theta_{Y}^2\rangle$, $\kappa_{T}$ decreases as the magnitude of angular displacements increases. Generally, higher temperature leads to more energetic internal motions with larger values of $\langle \theta_{i}^{2} \rangle$, so $\kappa_T$ should decrease with increasing $T$.  The observed NQR frequency $\nu\left( T \right) = \kappa_T \nu_0$ will then also decrease with increasing $T$ if $\nu_0$ does not increase with volume faster than $\kappa_T$.  

A net decrease in NQR frequency with increasing $T$ at atmospheric pressure is observed in most materials, though counter-examples exist \cite{kind1976cl} when either gradual or abrupt structural changes increase $\nu_0$ so sharply that the affect of $\kappa_T$ is overcome.  On the other hand, the volume variation of $\nu_0$ may sometimes be ignored, supposing that $\nu_0$ is a constant.  Whatever the dependence of $\nu_0$ on volume in a given material in fact is, $\nu_{0}(V)$ is readily accessible obtain using GIPAW calculations: simply prepare a set of crystal cells with lattice vectors and atomic positions each commensurate with the desired $T, V$, then, compute the static lattice NQR parameters $C_{q,0}$ and $\eta$ in each cell.

%
%
%
%
%
%
%
%
%
%
%
%
%

\section{Methods}

NQR frequencies depend on $T$ directly through internal motions and also depend on volume (and so $T$) as thermal expansion of the material causes structural changes that modulate all of the NQR parameters in potentially subtle ways.  In this study, both effects are accounted for by straight-forward computation of the volume dependence of $C_{q,0}$ and $\eta$, and the dependence of $\kappa_{T}$ on both $T$ and $V$.  The NQR frequency at a given $T, V$ is completely determined  by $C_{q,0}$, $\eta$, and $\kappa_T$ by equations \eqref{nqrspin32}, \eqref{nqrspin1}, and \eqref{kappaproduct}.  Accurate calculation of $\kappa_T$, $C_{q,0}$, and $\eta$ by a sequence of individual DFT calculations is the primary objective and result of this method.

\begin{figure}[ht] 
  \centering
      \includegraphics[width=8.6cm]{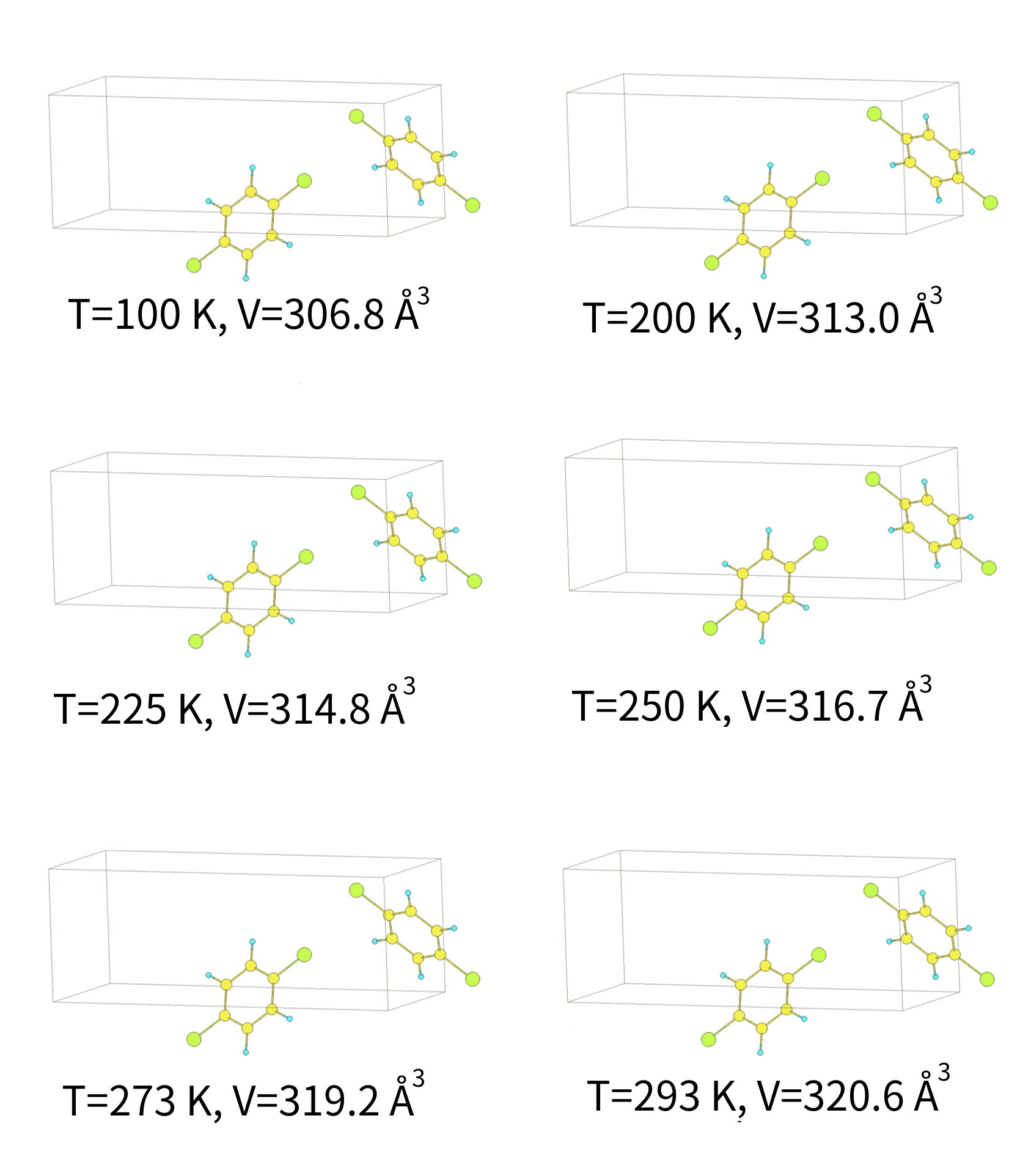}
        \caption{
                \label{fig:cell_boxes}  
                Schematic representation of the $\alpha$-pCl$_{2}\phi$ crystal cells generated to represent the structure at various temperatures.  An MD simulation and the accompanying sequence of EFG calculations for the time evolving system are carried out in each unit box, from which $C_{q,0}$ and $\kappa_T$ are extracted.  Each of the six crystal cell in the figure represents a single data point in the computed NQR temperature curve shown in figure \ref{fig:pcl2p_plot}.  To wit: computation of a singular data point \ref{fig:pcl2p_plot} requires an MD simulation of $N$ steps followed by $N$ SCF and $N$ EFG calculations.  In this case, $N=900$.
        }
\end{figure}

In the computation of the volume dependence of $C_{q,0}$ and $\eta$, available experimental data for atomic positions and lattice constants at various $T,V$ are used to prepare a set of "frozen" crystal cells, each with lattice vectors and equilibrium atomic positions intended to represent the system over a range of temperatures, as shown in figure \ref{fig:cell_boxes} .  In each of the prepared cells, the atomic positions are optimized and the EFG tensor components are computed using the GIPAW method in the optimized structure, yielding the full EFG tensor at each volume, the principle axes of which comprise the "space-fixed" coordinate system, against which the dynamical system's rotating EFG principle axes are compared in order to ascertain $\kappa_{T}$ as in equation \eqref{kappaT}.

Inclusion of the effect of lattice dynamics requires an MD simulation in each of the prepared cells, from which the evolution of the EFG axes in a system undergoing internal motion is computed. The MD simulation generates a set of atomic configurations in each of the optimized structures, intended to represent the system at the temperature which, according to experimental data, corresponds to the cell's volume at atmospheric pressure. The EFG tensors and the principle axes are computed in each of the perturbed structures generated by MD, and the rotation angles $\theta_X$, $\theta_Y$ which map the directions of the space-fixed axes to the updated axes are extracted.

\begin{figure}[ht] 
  \centering
      \includegraphics[width=8.6cm]{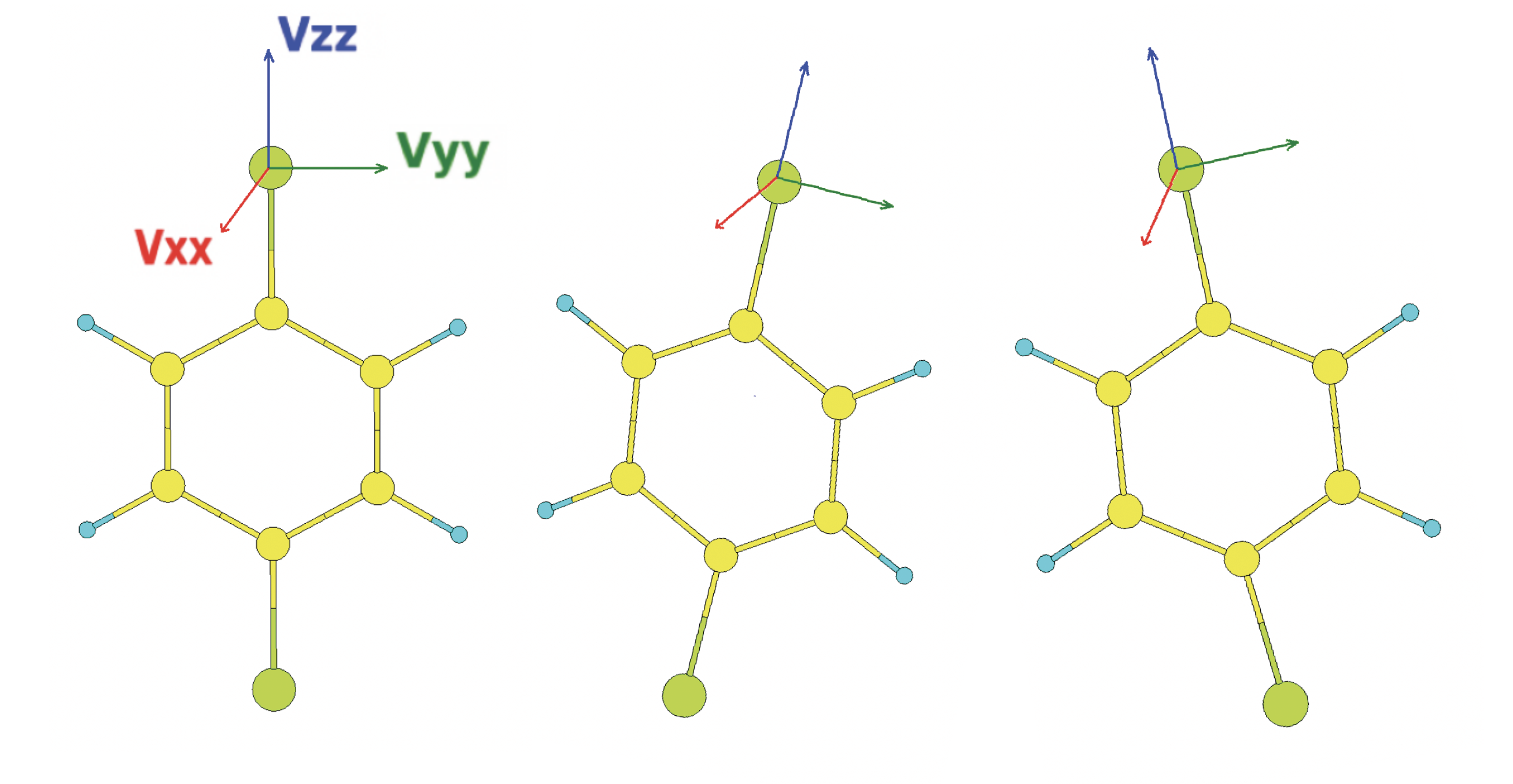}
        \caption{
                \label{fig:pcl2p_molecules}  
                Illustration of the fixed and moving principal axes of the EFG tensor at the position of chlorine in $\alpha$-pCl$_{2}\phi$. As the EFG at the $^{35}$Cl sites undergoes motions, the EFG components in the rotated system can be related to those in the fixed system through the angles $\theta_{X}$, $\theta_{Y}$, $\theta_{Z}$ developed by small intrinsic rotations about the $X^{\prime}$, $Y$, and $Z$ axes.  It is from the collection of these angles from an MD simulation that $\kappa_{T}$, the temperature correction to $\nu_{q}$ is computed at each volume.  A $T,V$ specific $\kappa_T$ must be obtained for each singular data point of calculated NQR frequencies shown in figures \ref{fig:pcl2p_plot} and \ref{fig:hmx_plot}.
                }
\end{figure}

Because intrinsic rotations do not commute we will, without loss of generality, transform the axes in the order $X^\prime-Y-Z$ to compute $\theta_{i}$.  The average of the squares of each $\theta_x$ and $\theta_y$ over the MD trajectory determine the dynamic normalization factor $\kappa_{T} = 1 - \frac32 \left(\langle \theta_{X}^2\rangle + \langle \theta_{Y}^2\rangle \right) - \frac12 \eta  \left(\langle \theta_{X}^2\rangle - \langle \theta_{Y}^2\rangle \right)$ which corrects the space-fixed $C_{q,0}$ for non-zero temperature.  As long as the $V,T$ pairs of the previous step be chosen to correspond to the experimentally determined crystal cell and atomic positions one would find in this temperature range at atmospheric pressure, the computed frequencies would represent the $T$-dependent, laboratory-observed NQR frequencies.

%
%
%

\section{Calculation Details}
All DFT calculations were carried out using the Quantum ESPRESSO software package with PBE psuedopotentials.  The geometry optimizations and the NVE molecular dynamics (MD) simulations were carried out with a cutoff energy of 60 Ry, and the EFG calculations were found to require a higher cutoff energy of 100 Ry.  For both the MD and EFG calculations, an SCF-convergence threshold of 10$^{-7}$ Ry was used. The MD integration time-step was made as long as possible in both systems by constraining some bond lengths and bond angles in each system.  In $\alpha$-paradichlorobenzene ($\alpha$-pCl$_{2}\phi$), 40 constraints were applied, leaving the chlorine's degrees of freedom as well as the translational and rotational degrees of freedom of the rigid benzene ring structure unconstrained.  In HMX 16 constraints were applied, freezing only the H-C bond-lengths.  The MD time-step for $\alpha$-pCl$_{2}\phi$ was 120 a.u. (5.8 fs) and for HMX 80 a.u. (3.9 fs).  Because chlorine is heavier than nitrogen, a longer simulation time was needed for $\alpha$-pCl$_{2}\phi$ than for HMX.  In both systems, the MD simulation was run for a total of 900 time steps, yielding 900 configurations used for the averaging of the EFG tensor in all cases.  In both systems, a Monkhorst-Pack K-mesh was used, 2x4x4 in the case of $\alpha$-pCl$_{2}\phi$, and 4x2x4 in the case of $\beta$-HMX.

%
%
%
%
%
%
%
%
%
%
%
%
\begin{table}[]
\footnotesize
\centering
\caption{The experimental cell parameters of $\alpha$-pCl$_{2}\phi$ \cite{wheeler1976intermolecular} \cite{estop1997alpha} and $\beta$-HMX \cite{deschamps2011thermal} used in this study.  The the variation of the lattice constants can, in general materials, either increase or decrease $C_{q,0}$.  In molecular crystals, increasing volume typically increases $C_{q,0}$, as intermolecular separation leads to greater anisotropy of the nuclear electronic environment.  Our results in table \ref{table:results-table} agree with this trend, making volume and temperature have competing effects on the observed NQR frequency, though the volume dependence of $C_{q,0}$ was found to be nearly an order of magnitude stronger in $\alpha$-pCl$_{2}\phi$ than in $\beta$-HMX as shown in figure \ref{fig:Cq0_vs_V} .\newline
}
\begin{tabular}{|c|c|c|c|c|c|c|} \hline\label{table:cell-parameters}
                             & T (K)  & a $(\ang)$     & b $(\ang)$      & c $(\ang)$     & $\beta$ & Volume $(\ang^3)$ \\ \hline
                             & 293 & 14.754 & 5.840   & 4.025  & 112.52$^\circ$  & 320.3  \\
                             & 273 & 14.747 & 5.830   & 4.014  & 112.33$^\circ$  & 319.2  \\
$\alpha$-pCl$_{2}\phi$ 	     & 250 & 14.730 & 5.812   & 3.997  & 112.24$^\circ$  & 316.7  \\
                             & 225 & 14.720 & 5.801   & 3.982  & 112.17$^\circ$  & 314.8  \\
                             & 200 & 14.705 & 5.787   & 3.967  & 112.00$^\circ$  & 311.4  \\
                             & 100 & 14.664 & 5.740   & 3.925  & 111.77$^\circ$  & 306.8  \\ \hline
                             & 303 & 6.5255 & 11.0369 & 7.3640 & 102.670$^\circ$ & 517.45 \\
                             & 293 & 6.5245 & 11.0240 & 7.3619 & 102.642$^\circ$ & 516.68 \\
                             & 273 & 6.5289 & 10.9875 & 7.3453 & 102.616$^\circ$ & 514.29 \\
                             & 248 & 6.5254 & 10.9702 & 7.3503 & 102.582$^\circ$ & 513.53 \\
$\beta$-HMX                  & 223 & 6.5334 & 10.9419 & 7.3421 & 102.491$^\circ$ & 512.45 \\
                             & 198 & 6.5206 & 10.9123 & 7.3395 & 102.467$^\circ$ & 509.93 \\
                             & 173 & 6.5273 & 10.8834 & 7.3286 & 102.366$^\circ$ & 508.54 \\
                             & 123 & 6.533  & 10.8400 & 7.3207 & 102.271$^\circ$ & 505.97 \\ \hline
\end{tabular}
\newline

\end{table}
%
%

%
%
%
%
\section{Structure Selection}
The present method is tested for two monoclinic molecular crystals $\alpha$-pCl$_{2}\phi$ and $\beta$-HMX.  These systems were chosen because of their relative structural simplicity as well as the great body of existing research on these systems, systems, to which the calculations could be compared for the purpose of validating the methods.   Each system is relatively simple, having one or few inequivalent nuclear sites and only two molecules in the unit cell.  Both systems contain relatively few atomic species and a small total number of atoms in the cell:  There are only 24 atoms in the unit cell of $\alpha$-pCl$_{2}\phi$ and the 56 atom $\beta$-HMX structure was the most tractable of the popular explosives (TNT, for example, contains 168 atoms in the cell).  Structural simplicity reduces  computation time, a convenience for convergence testing for the many DFT calculations required by the current method.  Performing the study on two quadrupolar nuclei $^{35}$Cl and $^{14}$N was also a motivation to choose one system containing each specie $^{35}$Cl and $^{14}$N.

Both systems have a long and extensive history of experimental results to which these calculations can be compared.  $\alpha$-pCl$_{2}\phi$ is the subject of some of the very earliest NQR research in history \cite{grechishkin1959nuclear} including on the temperature dependence of NQR \cite{vanier1960temperature} and was the subject of the seminal Bayer paper \cite{bayer1951theorie}.  The volume and stress dependence of its NQR frequencies has been characterized \cite{kushida1956dependence} \cite{zamar1988uniaxial}, and has persisted through time as a standard sample for testing new NQR techniques \cite{sullivan1971nuclear}.  The system's intermolecular interactions and dynamics, atomic positions in the crystal and dependence of the lattice vectors  on temperature have been well characterized. \cite{frasson1959structure}  \cite{wheeler1976intermolecular} \cite{estop1997alpha}

NQR frequencies in and internal motions in $\beta$-HMX have been studied extensively \cite{landers1981HMX} \cite{marino1982nitrogen} \cite{karpowicz1983librational} \cite{karpowicz1984comparison} \cite{buess2004factors}.  As an explosive material of great interest, there is also an abundance of existing computational work associated with HMX including simulations of its internal motions \cite{allis2006solid} \cite{sewell2003molecular}.  and recent DFT calculations of vibrational properties of $\beta$-HMX which have characterized the efficacy of local density approximation (LDA) vs the generalized gradient approximation (GGA) in the system. \cite{wu2011vibrational} The $T$-dependence of the lattice vectors has been determined\cite{deschamps2011thermal} and atomic positional data across a range of temperatures is available from the Cambridge Structural Database (CSD) in CIF form. \cite{groom2016cambridge}

\subsection{$\alpha$-paradichlorobenzene }

%
%

\begin{figure}[ht] 
  \centering
      \includegraphics[width=8.6cm]{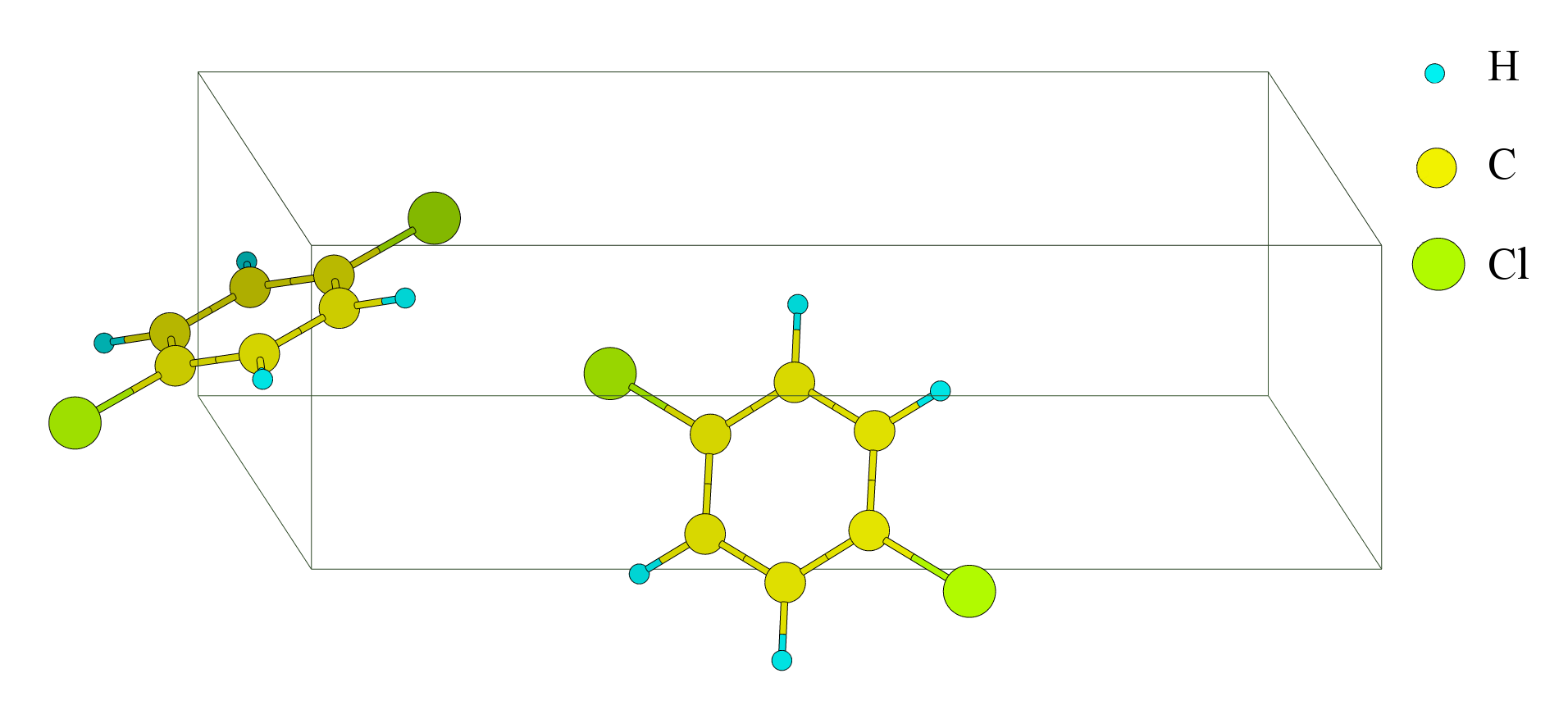}
        \caption{
                \label{fig:pcl2p_cell}  
                Monoclinic crystal structure of $\alpha$-paradichlorobenzene ($\alpha$-pCl$_{2}\phi$) at 293K \cite{estop1997alpha}.
        }
\end{figure}

Paradichlorobenzene has three polymorphs at room temperature.  In this study, we examine the alpha phase.  The cell is monoclinic, and contains two crystallographically equivalent molecules.  Each molecule contains two equivalent $^{35}$Cl sites, making only a single inequivalent $^{35}$Cl nucleus in this structure.  $^{35}$Cl has spin $\frac{3}{2}$, for which there is only one transition regardless of the magnitude of $\eta$. As a result, only a single NQR line is expected in this structure. \footnote{
We disregard the additional line one expects due to the heavy isotope $^{37}$Cl, which we did not account for here despite its natural abundance of 24.4\%. The NQR frequencies of the two isotopes should have the ratio of their respective quadrupole moments, $\nu_{^{35}\text{Cl}}/\nu_{^{37}\text{Cl}} = Q_{^{35}\text{Cl}}/Q_{^{37}\text{Cl}} = 1.2704$, under normal conditions where the nuclei are not subject to significant deformations.
}  Beginning with the structure determined by Estop et. al \cite{estop1997alpha}, which was specified at 293K, crystal cells corresponding to the known cell parameters at five lower temperatures were generated. To obtain the atomic positions at other temperatures, the experimental cell parameters \cite{wheeler1976intermolecular} were used to transform the atomic coordinates into the smaller cells at lower temperature.  This has the undesired effect of shrinking the bond lengths, so a full geometry optimization was performed after each incremental scaling.

%
%
\subsection{$\beta$-HMX}

\begin{figure}[ht] 
  \centering
      \includegraphics[width=8.6cm]{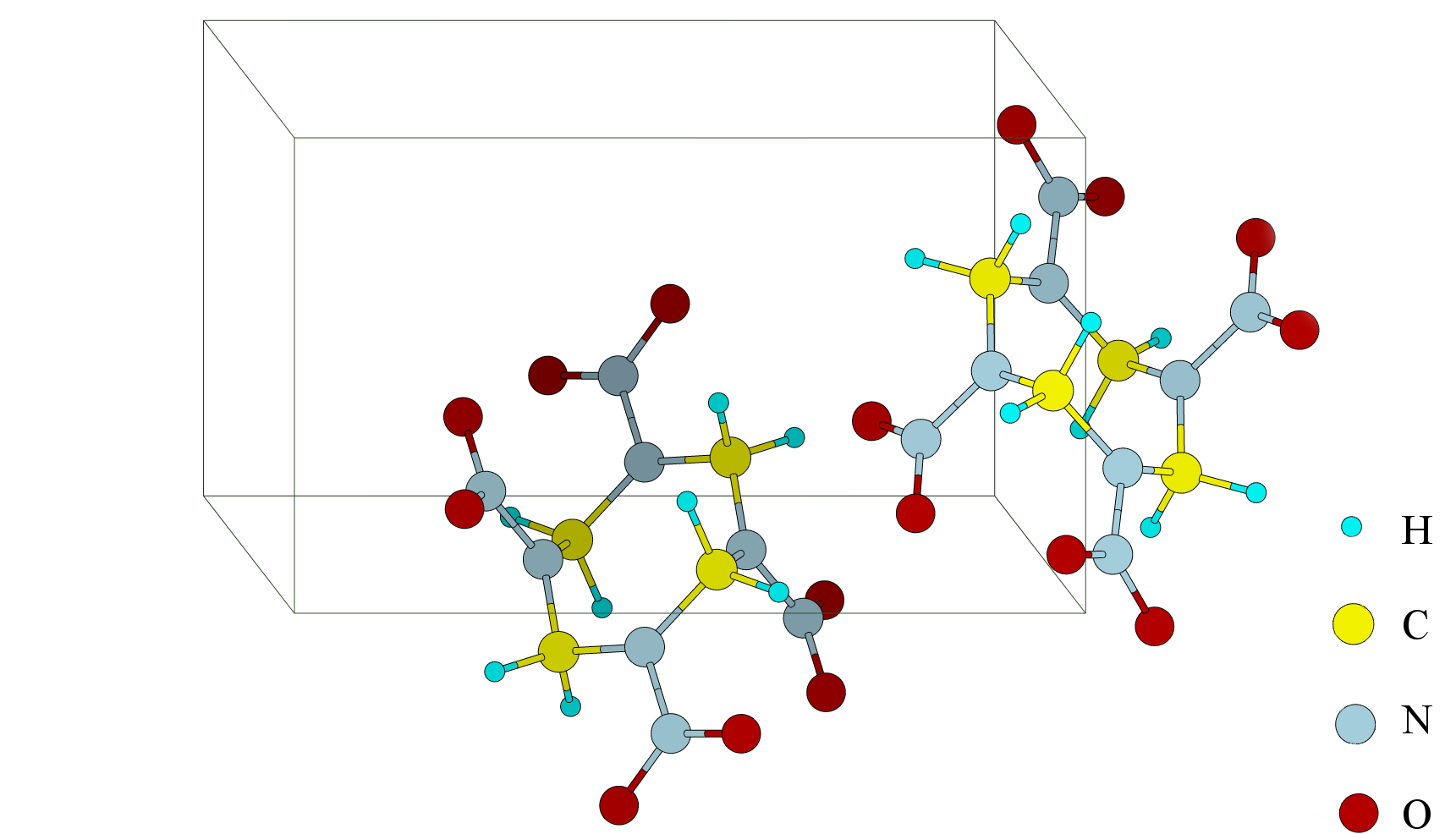}
        \caption{
                \label{fig:hmx_cell}  
                Monoclinic structure of $\beta$-HMX at 293K \cite{deschamps2011thermal}.
        }
\end{figure}

$\beta$-HMX is a monoclinic polymorph of HMX that is used as an explosives in military applications.  HMX has a a pseudo-octagonal ring structure with NO$_2$ substituents.  The structure has two crystallographically equivalent molecules in the cell.  Each molecule contains eight nitrogen sites, four of which are inequivalent.  The four nitrogens in each molecule belonging to NO$_2$ groups have very low NQR frequencies \cite{landers1981HMX}. The amine nitrogens, however have NQR frequencies above 5 MHz and are therefore of greater practical interest.  There are two inequivalent amine nitrogens in each molecule, differing most obviously by whether the NO$_{2}$ substituent is axial or equatorial, which are descriptors of the angle of inclination the N=N bond makes with the molecular pseudoplane.  Because $^{14}$N has spin 1, there are three NQR frequencies at each site in the case of an axially asymmetric field gradients.  The asymmetry parameters are found to be substantial (around 0.5) for both the axial and equatorial amine nitrogens.

%
%
%
%
%
%
%
\section{Results}


\begin{figure}[ht]
	\centering
	   	\includegraphics[width=8.6cm]{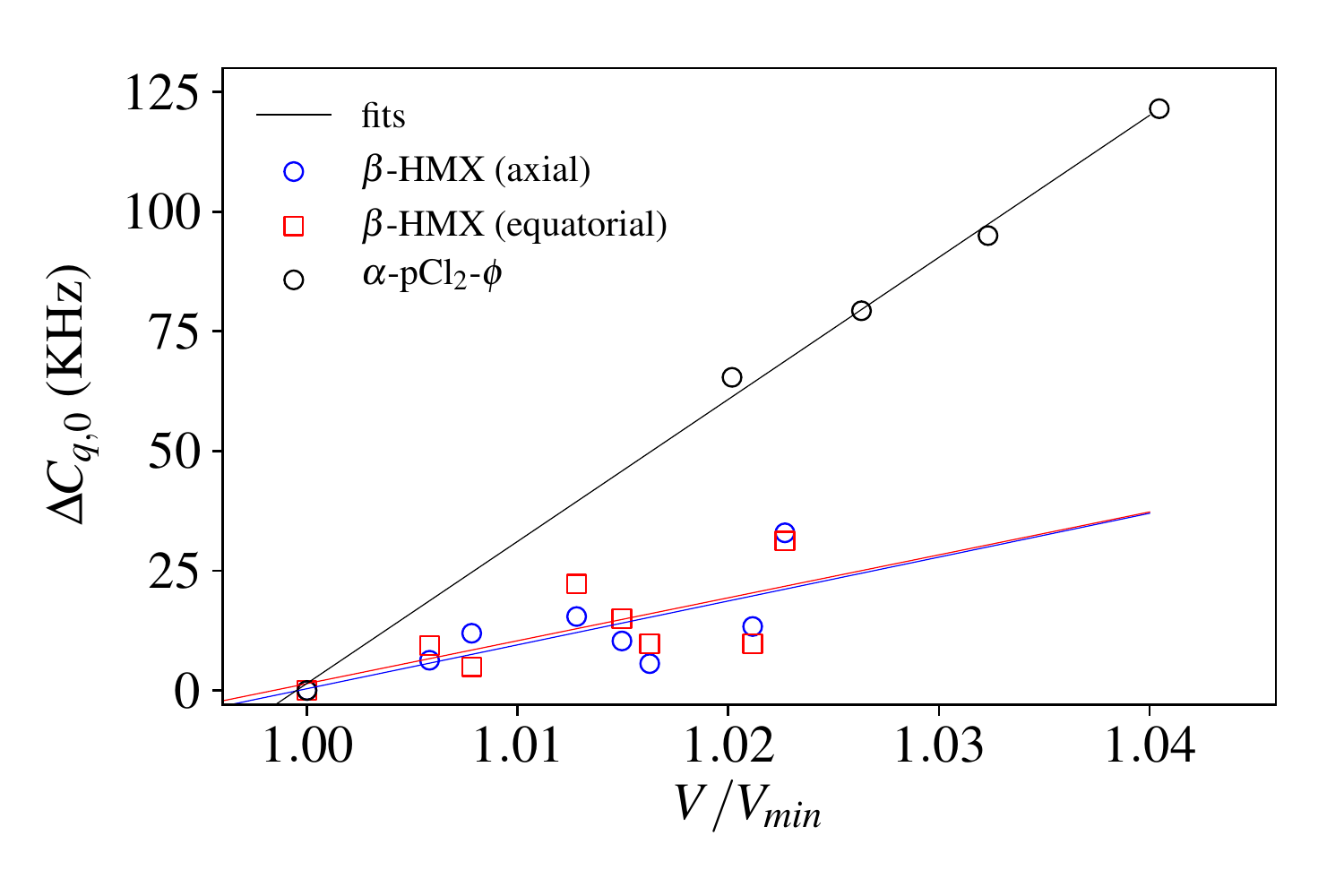}
		\vspace*{\floatsep}	
		\includegraphics[width=8.6cm]{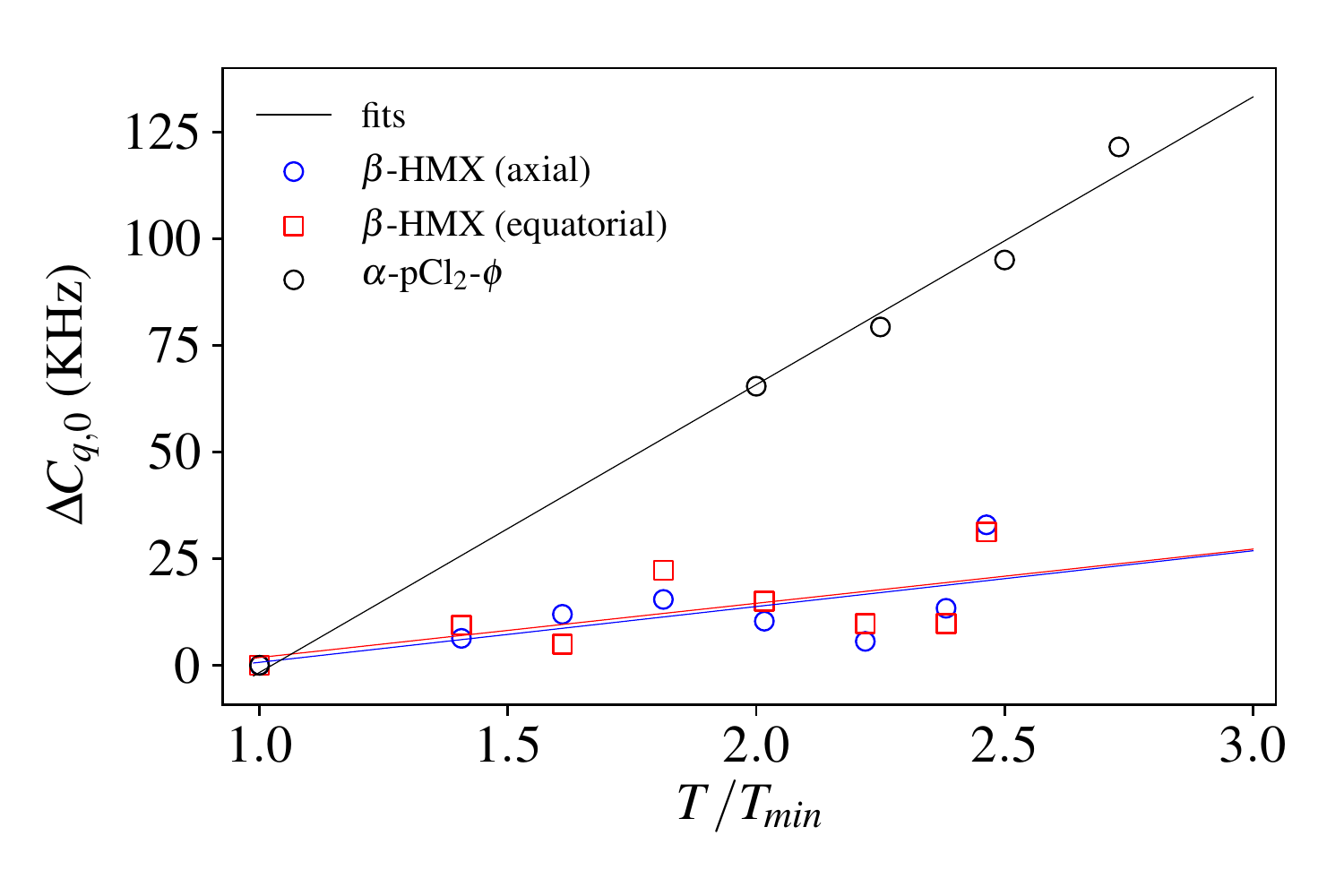}
  			\caption{
	               $C_{q,0}$ in $\alpha$-pCl$_{2}\phi$ exhibits a strong linear trend  with increasing volume that is not seen in $\beta$-HMX.  A strong intrinsic $T$-dependence (bottom figure) of $C_{q,0}$ on volume results due to thermal expansion in $\alpha$-pCl$_{2}\phi$ underpins the justification of use of a modified KBB fit in equation \eqref{modified-KBB}, which includes a term in $T^2$ by supposing that $C_{q,0}$ goes like $V$, and $V$ goes like $T$.  The $V$-dependence of $C_{q,0}$ is far smaller in $\beta$-HMX where effect can be ignored, and the use of the standard KBB function in equation \eqref{KBB-fit} is permitted.
			}
	   	\label{fig:Cq0_vs_V}
\end{figure}

The constant volume KBB model suggests a fitting function for $T$-dependence of the NQR frequencies of the form 

\begin{equation}\label{KBB-fit}
	\begin{aligned}
		\nu\left(T\right)  &= \nu_{0}\left(1 + a_{1}^{\prime}T + a_{-1}^{\prime}/T \right) \\
				   &= a_{0} + a_{1}T + a_{-1}/T   
	\end{aligned}
\end{equation}
where $\nu_{0} = a_{0}$ represents the static lattice NQR frequency and the other terms result from the internal motions which give rise to equation \eqref{tensors_thetas}.  However, because this model is a constant volume theory, it has long been understood \cite{wang1955pure} \cite{kushida1956dependence} \cite{das1958nuclear} the model may not suffice to describe the $T$-dependence of NQR frequencies outside of narrow temperature ranges because the static lattice coupling constant may vary significantly with cell volume. Furthermore, if the thermal expansion is anisotropic, as is the case in both systems under study in this work, variation of the cell parameters over temperature can impact $\eta$, resulting in a complicated implicit T-dependence of the parameter $\nu_{0} = a$ mediated by the thermal expansion.  In the case of $\beta$-HMX, the static lattice coupling constant was not found to depend strongly on the cell volume, and consequently, the standard KBB fitting function above was used for the fits in the case of HMX.  However, calculations presented in figure \ref{fig:Cq0_vs_V} show a strong linear dependence of $C_{q,0}$ on $V$ in $\alpha$-pCl$_{2}\phi$ that is not present in $\beta$-HMX, whose thermal expansion is not as significant to the local EFGs.  Examination of table \ref{table:cell-parameters} shows $V$ is linear in $T$ and calculations show that $\nu_{0}$ is linear in $V$; therefore, the supposition that $\nu_{0}$ is linear in T is justified and letting $\nu_{0} = a_{1}^{\prime\prime}T$ implies the appearance of a quadratic term $a_{2} T^2$ in equation \eqref{KBB-fit}

\begin{equation}\label{modified-KBB}
	\begin{aligned}
		\nu\left(T\right)  	&= \nu_{0}\left(1 + a_{1}^{\prime}T + a_{-1}^{\prime}/T \right) \\
					&= a_{1}^{\prime\prime}T\left(1 + a_{1}^{\prime}T + a_{-1}^{\prime}/T \right)\\
					&= a_{0} + a_{1}T + a_{2}T^2
	\end{aligned}
\end{equation}

Comparison of the absolute frequency predictions, as well as the fitting parameters arising from the experimental data and the calculations may help evaluate the success of this method in both of these structures.

\begin{figure}[ht]
	\centering
      			\includegraphics[width=8.6cm]{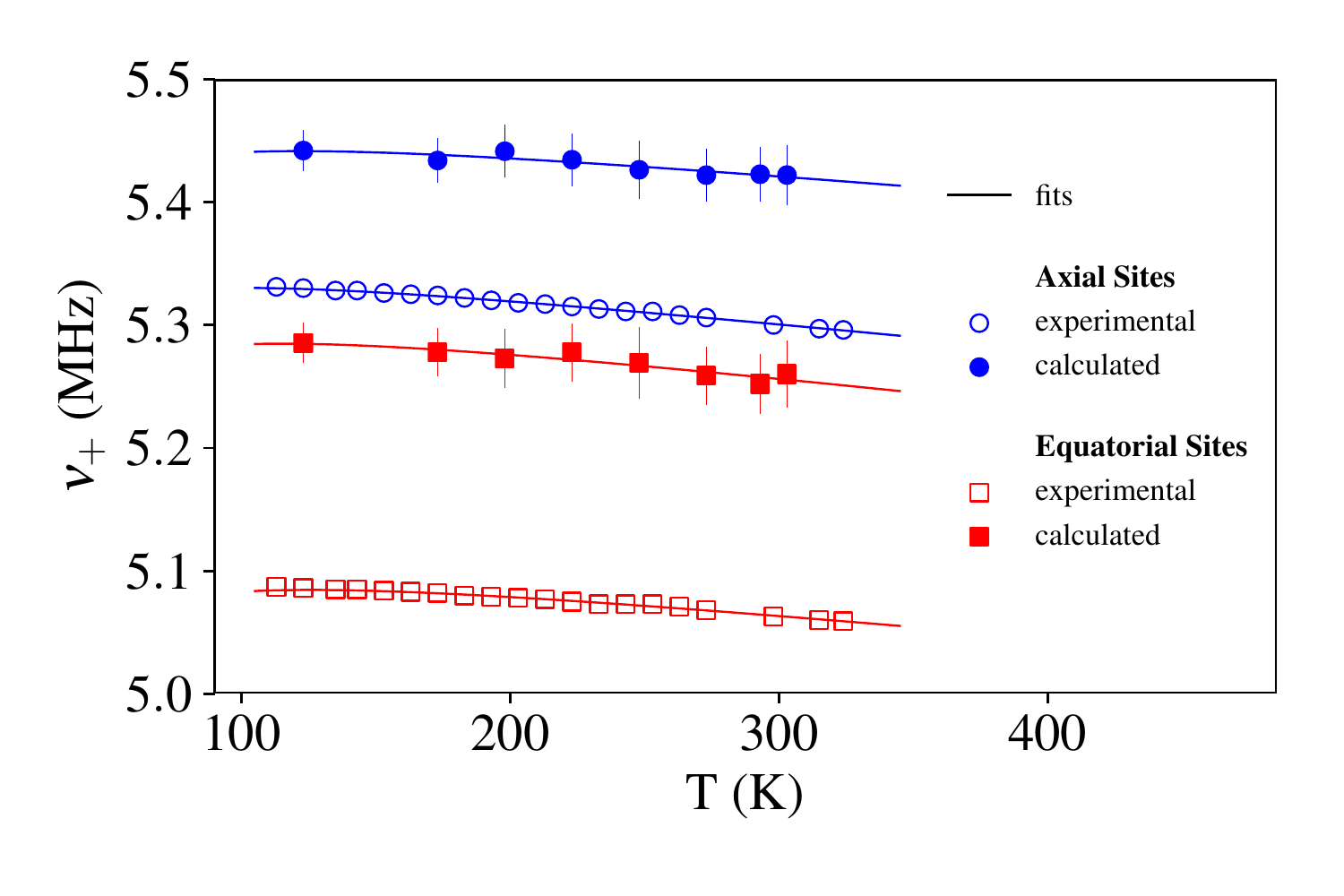}
       				\caption{
               			\label{fig:hmx_plot}  
The temperature dependence of the highest $^{14}$N NQR frequencies in amine sites of $\beta$-HMX calculated in this study (solid markers) compared with the experimental values (hollow markers). The circles denote the axial sites and the squares the equatorial sites.  The error bars are derived from the standard deviations of the sequences giving rise to $\langle C_{q,0} \rangle$ at each $T,V$. The calculations overestimate $C_{q,0}$ and the corresponding NQR frequency in both axial and equatorial sites, with error percentages for $a_0$ at 4\%, and 2\% for equatorial and axial sites, respectively as shown in table \ref{table:fits2-table}. }
\end{figure}

\subsection{$\beta$-HMX}

The calculations systematically overestimate the NQR frequency in both axial and equatorial sites, likely due to error in the determination of $C_{q,0}$ since the magnitude of $\phi_{ZZ}$, defined as the largest eigenvalue of the EFG tensor, is highly sensitive to the DFT simulation parameters chosen. In particular, the suitability and quality of the pseudopotentials utilized to compute the charge densities were found to significantly impact the computed EFG tensors, even when the cutoff energies and other simulation parameters are converged independently.  The degree of absolute error is reflected directly by the fitting term $a_0$. shown in table \ref{table:fits2-table}.  In a static lattice, small changes in the atomic positions can also affect $C_{q,0}$ significantly; however, in this case, error introduced by atomic positional discrepancy is reduced by averaging $C_{q,0}$ over the MD trajectory. In principle one could obtain $C_{q,0}$ by performing a very well converged geometry optimization and computing the EFGs in the output.  The equatorial calculations overestimate the NQR frequencies by twice as much as the axial calculations, yielding error percentages in the constant fitting parameter $a_0$ at 4\%, and 2\% for equatorial and axial sites, respectively.  Consequently, the calculations underestimate the absolute difference between the axial and equatorial NQR frequencies by about a factor of $\frac12$.  The electronic environments of the two sites types differ primarily due to the orientation of the substituent NO$_{2}$ molecule with respect to the molecular pseudo-plane, to which the DFT is sensitive. Using the fitting function $\nu\left(T\right) = a_{0} + a_{1}T + a_{-1}/T $ the experimental data from \cite{landers1981HMX} for showed the axial amines to have generally NQR frequencies about 200kHz higher than the equatorial sites, and a slightly sharper fall-off with temperature.  No additional quadratic term was used to fit either experimental or calculated frequencies because the computed static lattice values of $C_{q,0}$ did not strictly increase with increasing volume, as shown in figure \ref{fig:Cq0_vs_V}.


%
%
%

\begin{figure}
	\centering

      			\includegraphics[width=8.6cm]{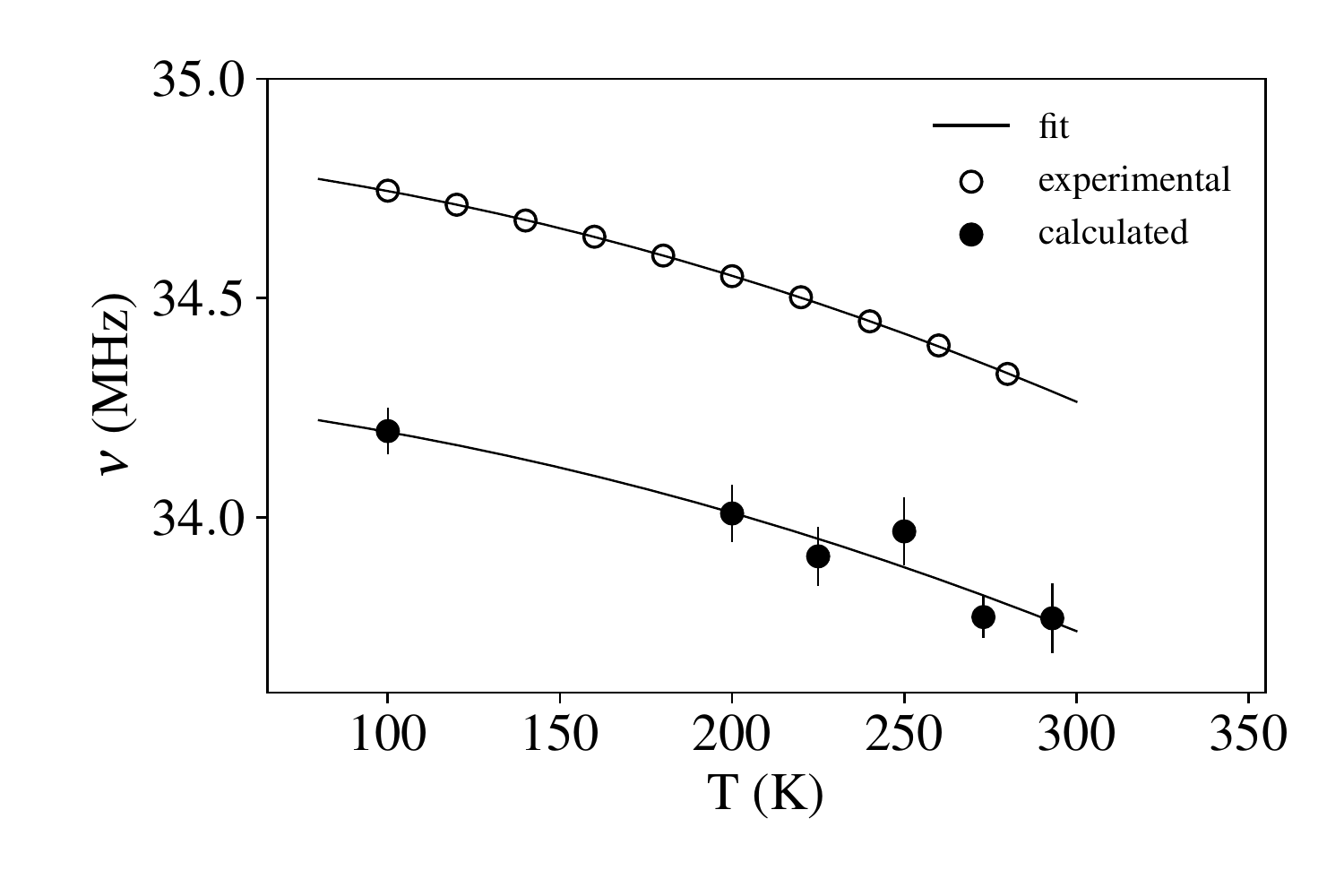}
        		\caption{
               		\label{fig:pcl2p_plot}  
The temperature dependence of the NQR frequency in $\alpha$-pCl$_{2}\phi$ calculated in this study (solid circles) compared with the experimental values (hollow circles).  The fit was found to require a quadratic term as in equation \eqref{modified-KBB} due to the influence of thermal expansion on the static lattice EFGs presented in figure \ref{fig:Cq0_vs_V}. The constant and quadratic fitting parameters of the calculated frequencies are within 1.6\%  and 12\% of the experimental fit, respectively. The error bars show are derived from the standard deviations of the sequences of values of $C_{q,0}$ computed over the molecular dynamics simulation outputs, being the primary averaged quantity associated with each NQR frequency.  }
\end{figure}

\subsection{$\alpha$-paradichlorobenzene}

In this structure, the calculations systematically underestimate the NQR frequency by about 500KHz, as shown in figure \ref{fig:pcl2p_plot}.  The cause of the underestimation is again likely error in the estimation of $C_{q,0}$ which has a large impact on the absolute errors in the calculation.  The T-dependence of NQR frequencies $\alpha$-paradichlorobenzene was fit using a modified KBB fitting function given by equation \eqref{modified-KBB}, $\nu\left(T\right) = a_0+a_{1}T+a_{2}T^2$.  The experimental NQR data resulting in fitting parameters $a_{0}=34.8437$  MHz, $a_{1}=-5.2424\times10^{-4} \text{ MHz} \cdot T^{-1}$, and  $a_{2}=-4.6970\times 10^{-6} \text{ MHz}\cdot T^{-2}$. The corresponding fitting parameters $a_{i}$ from the fits of the calculated NQR frequencies are $a_{0}=34.2932$  MHz, $a_{1}=-5.5559\times10^{-4} \text{ MHz} \cdot T^{-1}$, and  $a_{2}=-4.2894\times 10^{-6} \text{ MHz}\cdot T^{-2}$. 

The fitting parameters are summarized in table \ref{table:fits2-table}.  The constant terms agree to within 1.6\% and the dominant quadratic terms agree within 12\%, while the linear terms differ by 26\%.  The absolute error of the NQR frequency prediction across the  100K-293K temperature range was at most 580 KHz and at its lowest 450KHz, while the amount by which the measured NQR frequency drifts over this temperature range is similarly about 450 KHz.

%
%

\begin{figure}[ht] 
  \centering
      \includegraphics[width=8.6cm]{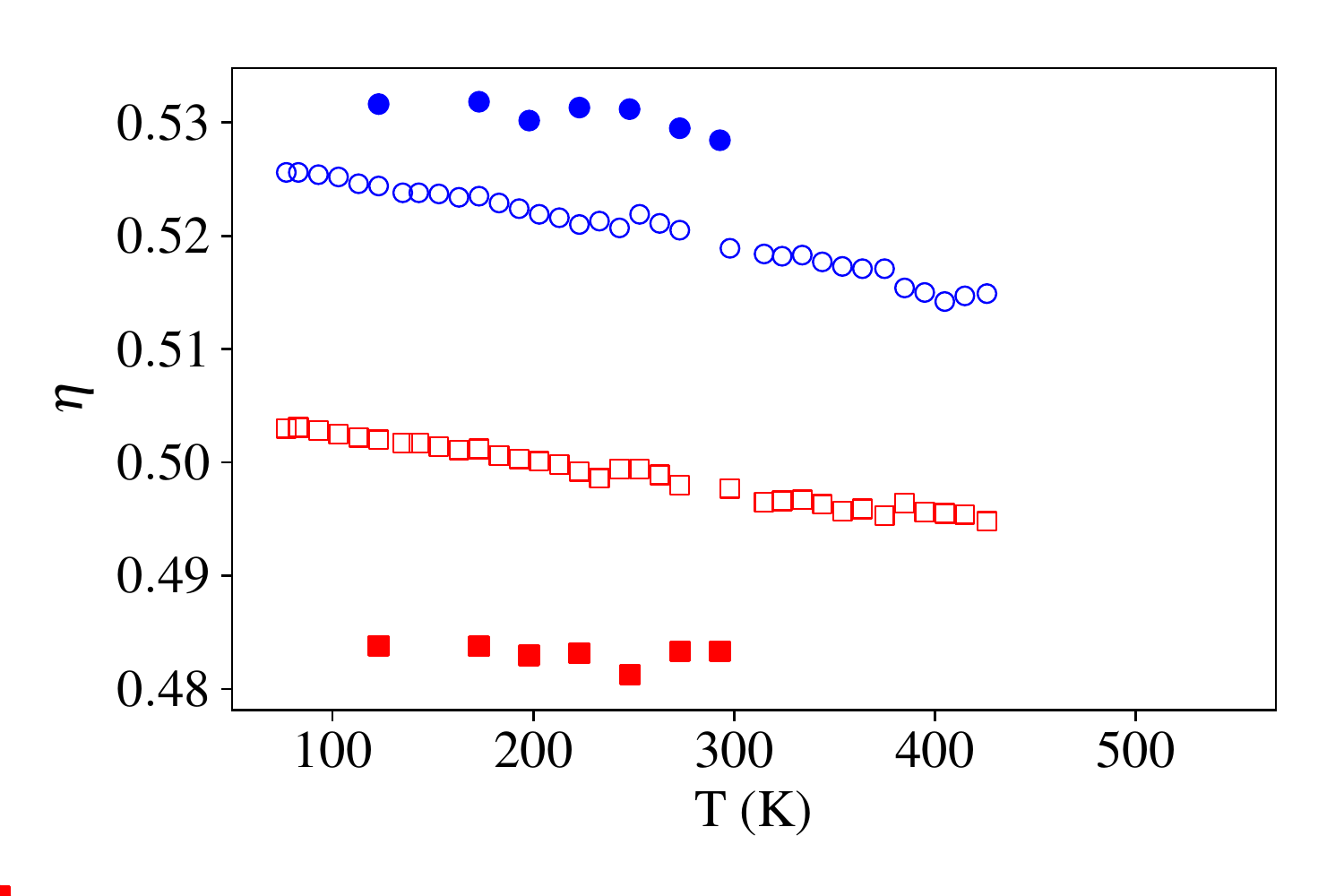}
        \caption{
                \label{fig:eta_plot}  
                Comparison of experimental and calculated asymmetry parameters for the amine nitrogen sites in $\beta$-HMX.  The calculated $\eta$ shown are averaged over the equivalent sites and the MD simulation steps.  Unlike the $C_{q,0}$ whose volume dependence is reliably positive for molecular crystals, the asymmetry parameter can, in general materials, display any trend with increasing volume. Despite this, the calculations captured the trend reasonably well, though underestimating the magnitude of the slope in the equatorial case.  The calculations overestimate the axial $\eta$ and underestimate the equatorial $\eta$; notably, the ratio of the associated absolute errors is roughly the same for the NQR frequencies displayed in \ref{fig:hmx_plot}, being nearly equal to $\frac12$ and independent of $T,V$.
}
\end{figure}

%
%

\begin{table*}[]
	\footnotesize
	\centering
	
	\caption{Comparison of the computed T-dependent NQR parameters from this work with experimental values in $\alpha$-pCl$_{2}\phi$ \cite{moross1966temperature} and $\beta$-HMX.  \cite{landers1981HMX}   The parameter $C_{q,0}$ is derived as an average of the $C_{q,0}^{i}$ computed with the $i^{th}$ output of the molecular dynamics simulation. $\kappa_T$ is given by equation \eqref{kappaT}.  Note that mean values of $C_{q,0}=6.11$ MHz have been imputed for HMX's equatorial sites at 273K and 293K because the computed averages showed signs of systematic error.  The absolute error column contains the absolute value of the difference between the calculated and experimental NQR frequency $\nu(T)$.\newline
	}

	\tabcolsep=0.1cm
	\begin{tabular}{|c|c|c|c|c|c|c|c|c|c|c||c}													\hline
		& \multicolumn{1}{l|}{} & \multicolumn{5}{c|}{\textbf{computed NQR parameters}} & \multicolumn{1}{c|}{\textbf{experimental}} & \multicolumn{1}{c|}{\textbf{error}}                                     \\
		\textbf{structure}     & T  (K) & $C_{q,0}$ (MHz) & $\eta_0$ & $\nu_0$ (MHz) & $\kappa_{T}$ & $\nu \left(T\right)$ (MHz) & $\nu \left( T \right)$ (MHz) & $ \Delta \nu$ (MHz) \\ \hline


		                       	& 293 & 69.3100 & 0.09749 & 34.7099 & 0.9729 & 33.7696 & 34.2869 & 0.5486	\\
					& 273 & 69.3241 & 0.09714 & 34.7165 & 0.9728 & 33.7725 & 34.3505 & 0.5424	\\		
		$\alpha$-pCl$_{2}\phi$ 	& 250 & 69.2976 & 0.09687 & 34.7029 & 0.9788 & 33.9679 & 34.4190 & 0.5771	\\
		                       	& 225 & 69.2819 & 0.09650 & 34.6947 & 0.9774 & 33.9109 & 34.4880 & 0.4511	\\
					& 200 & 69.2680 & 0.09635 & 34.6875 & 0.9804 & 34.0088 & 34.5511 & 0.5780 	\\
					& 100 & 69.2026 & 0.09566 & 34.6540 & 0.9868 & 34.1966 & 34.7452 & 0.5173 	\\ \hline


		                       	& 303 & 6.2343 & 0.5312 & 5.5036 & 0.9852 & 5.4219 & 5.3008 & 0.1109	\\
					& 293 & 6.2147 & 0.5284 & 5.4821 & 0.9891 & 5.4226 & 5.3026 & 0.1105	\\
					& 273 & 6.2070 & 0.5295 & 5.4769 & 0.9900 & 5.4219 & 5.3061 & 0.1221 	\\
		$\beta$-HMX            	& 248 & 6.2117 & 0.5312 & 5.4837 & 0.9895 & 5.4262 & 5.3105 & 0.1195  	\\
		(axial sites)          	& 223 & 6.2168 & 0.5313 & 5.4884 & 0.9902 & 5.4344 & 5.3149 & 0.1156  	\\
		                      	& 198 & 6.2133 & 0.5302 & 5.4835 & 0.9923 & 5.4413 & 5.3191 & 0.1157  	\\
		                       	& 173 & 6.2077 & 0.5318 & 5.4811 & 0.9914 & 5.4337 & 5.3233 & 0.1200	\\ \hline


		              	        & 303 & 6.1314 & 0.4850 & 5.3420 & 0.9847 & 5.2601 & 5.0643 & 0.1984	\\
					& 293 & 6.1100 & 0.4833 & 5.2642 & 0.9871 & 5.2521 & 5.0657 & 0.1964  	\\
					& 273 & 6.1100 & 0.4833 & 5.2593 & 0.9883 & 5.2587 & 5.0685 & 0.1945	\\
		$\beta$-HMX            	& 248 & 6.1152 & 0.4813 & 5.3222 & 0.9901 & 5.2693 & 5.0718 & 0.2026 	\\
		(equatorial sites)     	& 223 & 6.1225 & 0.4832 & 5.3314 & 0.9899 & 5.2779 & 5.0751 & 0.1974 	\\
					& 198 & 6.1052 & 0.4830 & 5.3160 & 0.9919 & 5.2729 & 5.0784 & 0.1903 	\\
					& 173 & 6.1096 & 0.4838 & 5.3211 & 0.9919 & 5.2778 & 5.0815 & 0.1864 	\\
					& 123 & 6.1002 & 0.4838 & 5.3130 & 0.9948 & 5.2854 & 5.0869 & 0.1957 	\\ \hline
	\end{tabular}
	 
	\label{table:results-table}
\end{table*}

\section{Discussion}

\subsection{Advantages and Disadvantages}
The method proposed has the advantage that it can be applied in a straightforward fashion to any system in which data for atomic positions and the thermal expansion of the lattice vectors are available, without the need for additional postulates or supporting experimental data such as IR or Raman studies.  A disadvantage of the current application of the method is that the computational cost can be high;  in the present study, an MD simulation of 900 steps followed by 900 SCF/EFG calculations were used to generate a single data point on the plots in figures \ref{fig:hmx_plot} and \ref{fig:pcl2p_plot}.  Six and sixteen data points were computed in $\alpha$-pCl$_{2}\phi$ and $\beta$-HMX, respectively.

\begin{table*}[]
	\footnotesize
	\caption{Fitting parameters of calculated and experimental NQR frequencies for functional form $\nu\left(T\right)=a_0 + a_1 T + a_2 T^2 + a_{-1}/T$. \newline}
	\tabcolsep=0.1cm
	\begin{center}
		\begin{tabular}{| l  r || c | c | c | c | }
			\hline

\textbf{structure}  	& \textbf{dataset} 	& \multicolumn{4}{c|}{\textbf{fitting parameters}}          \\ 
			&			& $a_0 $ (MHz) 	& $a_1 $  $\left(\frac{\text{MHz}\cdot\text{K}^{-1}}{10^{-4}}\right)$ & $a_2 $ $\left(\frac{\text{MHz}\cdot\text{K}^{-2}}{10^{-6}}\right)$ 					       & $a_{-1} \left(\text{MHz}\cdot\text{K}\right)$ \\ \hline 
			             
						&	experimental 		& 34.8437 & -5.2424 & -4.6970 	& ---     \\
$\alpha$-pCl$_{2}$-$\phi$				&	calculated   		& 34.2932 & -5.5595 & -4.2895 	& ---     \\
						&	error (no units)	& 1.56\%  & 25.8\%  & 12.9\%    & ---     \\   \hline
						&	experimental 		& 5.3761  & -2.2642 & ---       & -2.3392 \\
$\beta$-HMX (axial)					&	calculated   		& 5.4885  & -1.9402 & ---       & -2.8598 \\
						&	error (no units)	& 2.09\%  & 14.3\%  & ---       & 22.3\%  \\ \hline
						&	experimental 		& 5.1372  & -2.0976 & ---       & -3.3160 \\
$\beta$-HMX (equatorial)				&	calculated   		& 5.3426  & -2.5139 & ---       & -3.3235 \\
						&	error (no units) 	& 4.00\%  & 19.84\% & ---       & 0.226\% \\
			
			\hline
			
		\end{tabular}
	\end{center}
	
	\label{table:fits2-table}
\end{table*}

%
%

\subsection{Considerations for Proper Implementation}
The technique presented here is structure agnostic because the heavy-lifting is done by the DFT calculations.  In principle, the results will be accurate in other systems provided the MD simulations faithfully represent the structure's dynamics at the desired temperatures, and the subsequent sets of EFG calculations are carried out with sufficient simulation parameters for accuracy, requiring some care in the application of the simulations.  Attention to detail regarding both the underlying physics and best practices in computing is essential for a successful implementation, especially for the management of the output data, having both considerable size and complexity.

For the EFG calculations, careful choice of pseudopotentials and the simulation parameters of the DFT calculations, such as cutoff energies and SCF convergence thresholds is needed in order to ensure convergence of the EFGs.  It is also worth considering the precision of the known quadrupole moments in the nuclei under study, and recognition that the values of $Q$ used in the EFG calculations are constant here, though deformation of the nucleus by the surroundings can occur and effectively change the instantaneous value of $Q$ for the nucleus. \footnote{NQR measurements involve direct observation of the coupling constants $eQ\phi_{ZZ}/h$, whereas in the present study, we leave  $Q$ a constant and account for dynamical effects on $\phi_{ZZ}$. Values of $Q$ in standard tables are of the intrinsic quadrupole moments.  But the nuclear environment may cause deformations of the nucleus that induce contributions to the quadrupole moment.} Finally, One must perform the summary operation of averaging the EFG tensors in the correct manner to compute the correction to the NQR frequency computed in the static lattice.

The data generated may have considerable size on disk making computation of the $\langle\theta_{i}\rangle$ cumbersome without some consideration taken in the efficient storage and processing of the default simulation outputs. On occasion, the convention $|\phi_{ZZ}| \ge |\phi_{YY}| \ge |\phi_{XX}|$ will cause a sudden swap in the names EFG's eigenvectors if the magnitude of $\phi_{XX}$ momentarily overtakes $\phi_{YY}$, leading to a discontinuity in the time series of the $\theta_{i}$.  This can be remedied by checking that the moving principal axes coordinate system remains right handed.

\section{Conclusion}

The calculations underestimate the NQR frequencies in $\alpha$-pCl$_{2}\phi$ and overestimate them in $\beta$-HMX.  The overestimation is about twice are large in the equatorial sites than for the axial ones.  The calculated and experimental fits agreed reasonably well when the experimental data was restricted to the same temperature range covered by the calculations.  In $\alpha$-pCl$_{2}\phi$ the fitting of the calculated NQR frequencies match the experimental fits to 1.56\%, and 12.9\% for the constant $a_{0}$ and quadratic $a_{2}$ terms respectively, but the linear term $a_{1}$ displayed error in excess of 25\%.

In $\beta$-HMX the experimental and calculated constants $a_{0}$ agreed to 2.09\% and 4.00\% for axial and equatorial sites, respectively.  The errors in the linear terms were higher at 14.3\% and 19.9\%.  The $1/T$ fitting terms $a_{-1}$ agreed to less than 0.5\% in the equatorial case but the error was 22.3\% in the axial case.  The $a_{-1}$ term likely requires data at lower temperatures to be fit reliably, as the lowest temperature calculated in $\beta$-HMX was 123K. 

The method presented may be applied to arbitrary solids but more work is needed to extend the method to larger systems in a scalable manner.  Repeated application of the general method is needed to confirm its robustness, for which automation of both the input data preparation and output data processing would be of great use since the datasets generated can reach an unruly size on disk in systems of even modest size, without clever implementation.  On-the-fly data pruning, parsing and processing will likely be needed to extend this method to very large systems.  

The DFT computations carried out in this procedure produce a rich dataset that may be exploited in various ways that may lend improved results or computational efficiency, such as in the manner of tensor averaging used to compute $\kappa_{T}$.  There is also the potential to train statistical models to map atomic positions to EFGs which, if effective, would substantially increase efficiency as the molecular dynamics simulations are significantly cheaper than the SCF and EFG computations in terms of CPU time.

%
%
 
\section{Acknowledgements}
This work is supported in part by the National Science Foundation grant NSF-DMR-1157490 and Department of Energy grants DE-FG02-02ER45995. The authors extend gratitude to NERSC for computational resources and its support staff for assistance with compilations and automation procedures. 

\bibliography{main} 

\begin{thebibliography}{62}%
\makeatletter
\providecommand \@ifxundefined [1]{%
 \@ifx{#1\undefined}
}%
\providecommand \@ifnum [1]{%
 \ifnum #1\expandafter \@firstoftwo
 \else \expandafter \@secondoftwo
 \fi
}%
\providecommand \@ifx [1]{%
 \ifx #1\expandafter \@firstoftwo
 \else \expandafter \@secondoftwo
 \fi
}%
\providecommand \natexlab [1]{#1}%
\providecommand \enquote  [1]{``#1''}%
\providecommand \bibnamefont  [1]{#1}%
\providecommand \bibfnamefont [1]{#1}%
\providecommand \citenamefont [1]{#1}%
\providecommand \href@noop [0]{\@secondoftwo}%
\providecommand \href [0]{\begingroup \@sanitize@url \@href}%
\providecommand \@href[1]{\@@startlink{#1}\@@href}%
\providecommand \@@href[1]{\endgroup#1\@@endlink}%
\providecommand \@sanitize@url [0]{\catcode `\\12\catcode `\$12\catcode
  `\&12\catcode `\#12\catcode `\^12\catcode `\_12\catcode `\%12\relax}%
\providecommand \@@startlink[1]{}%
\providecommand \@@endlink[0]{}%
\providecommand \url  [0]{\begingroup\@sanitize@url \@url }%
\providecommand \@url [1]{\endgroup\@href {#1}{\urlprefix }}%
\providecommand \urlprefix  [0]{URL }%
\providecommand \Eprint [0]{\href }%
\providecommand \doibase [0]{http://dx.doi.org/}%
\providecommand \selectlanguage [0]{\@gobble}%
\providecommand \bibinfo  [0]{\@secondoftwo}%
\providecommand \bibfield  [0]{\@secondoftwo}%
\providecommand \translation [1]{[#1]}%
\providecommand \BibitemOpen [0]{}%
\providecommand \bibitemStop [0]{}%
\providecommand \bibitemNoStop [0]{.\EOS\space}%
\providecommand \EOS [0]{\spacefactor3000\relax}%
\providecommand \BibitemShut  [1]{\csname bibitem#1\endcsname}%
\let\auto@bib@innerbib\@empty
\bibitem [{\citenamefont {Pound}(1950)}]{pound1950nuclear}%
  \BibitemOpen
  \bibfield  {author} {\bibinfo {author} {\bibfnamefont {R.}~\bibnamefont
  {Pound}},\ }\href {\doibase 10.1103/PhysRev.79.685} {\bibfield  {journal}
  {\bibinfo  {journal} {Phys. Rev.}\ }\textbf {\bibinfo {volume} {79}},\
  \bibinfo {pages} {685} (\bibinfo {year} {1950})}\BibitemShut {NoStop}%
\bibitem [{\citenamefont {Dodgen}\ and\ \citenamefont
  {Ragle}(1956)}]{dodgen1956observation}%
  \BibitemOpen
  \bibfield  {author} {\bibinfo {author} {\bibfnamefont {H.}~\bibnamefont
  {Dodgen}}\ and\ \bibinfo {author} {\bibfnamefont {J.}~\bibnamefont {Ragle}},\
  }\href {\doibase 10.1063/1.1742915} {\bibfield  {journal} {\bibinfo
  {journal} {J. Chem. Phys.}\ }\textbf {\bibinfo {volume} {25}},\ \bibinfo
  {pages} {376} (\bibinfo {year} {1956})}\BibitemShut {NoStop}%
\bibitem [{\citenamefont {Kind}\ and\ \citenamefont {Roos}(1976)}]{kind1976cl}%
  \BibitemOpen
  \bibfield  {author} {\bibinfo {author} {\bibfnamefont {R.}~\bibnamefont
  {Kind}}\ and\ \bibinfo {author} {\bibfnamefont {J.}~\bibnamefont {Roos}},\
  }\href {\doibase 10.1103/PhysRevB.13.45} {\bibfield  {journal} {\bibinfo
  {journal} {Phys. Rev. B}\ }\textbf {\bibinfo {volume} {13}},\ \bibinfo
  {pages} {45} (\bibinfo {year} {1976})}\BibitemShut {NoStop}%
\bibitem [{\citenamefont {Karpowicz}\ and\ \citenamefont
  {Brill}(1983)}]{karpowicz1983librational}%
  \BibitemOpen
  \bibfield  {author} {\bibinfo {author} {\bibfnamefont {R.}~\bibnamefont
  {Karpowicz}}\ and\ \bibinfo {author} {\bibfnamefont {T.}~\bibnamefont
  {Brill}},\ }\href {\doibase 10.1021/j100235a017} {\bibfield  {journal}
  {\bibinfo  {journal} {J. of Phys. Chem.}\ }\textbf {\bibinfo {volume} {87}},\
  \bibinfo {pages} {2109} (\bibinfo {year} {1983})}\BibitemShut {NoStop}%
\bibitem [{\citenamefont {Karpowicz}\ and\ \citenamefont
  {Brill}(1984)}]{karpowicz1984comparison}%
  \BibitemOpen
  \bibfield  {author} {\bibinfo {author} {\bibfnamefont {R.~J.}\ \bibnamefont
  {Karpowicz}}\ and\ \bibinfo {author} {\bibfnamefont {T.~B.}\ \bibnamefont
  {Brill}},\ }\href {\doibase 10.1021/j150647a005} {\bibfield  {journal}
  {\bibinfo  {journal} {J. Phys. Chem.}\ }\textbf {\bibinfo {volume} {88}},\
  \bibinfo {pages} {348} (\bibinfo {year} {1984})}\BibitemShut {NoStop}%
\bibitem [{\citenamefont {Das}\ and\ \citenamefont
  {Hahn}(1958)}]{das1958nuclear}%
  \BibitemOpen
  \bibfield  {author} {\bibinfo {author} {\bibfnamefont {T.~P.}\ \bibnamefont
  {Das}}\ and\ \bibinfo {author} {\bibfnamefont {E.~L.}\ \bibnamefont {Hahn}},\
  }\href@noop {} {\emph {\bibinfo {title} {Nuclear Quadrupole Resonance
  Spectroscopy}}},\ \bibinfo {number} {1}\ (\bibinfo  {publisher} {Academic
  Pr},\ \bibinfo {year} {1958})\BibitemShut {NoStop}%
\bibitem [{\citenamefont {Dass}\ and\ \citenamefont
  {Gajjar}(2011)}]{dass2011anti}%
  \BibitemOpen
  \bibfield  {author} {\bibinfo {author} {\bibfnamefont {R.}~\bibnamefont
  {Dass}}\ and\ \bibinfo {author} {\bibfnamefont {B.}~\bibnamefont {Gajjar}},\
  }\href {\doibase 10.4018/ijudh.2011100104} {\bibfield  {journal} {\bibinfo
  {journal} {Intern. J.of User-Driven Healthcare (IJUDH)}\ }\textbf {\bibinfo
  {volume} {1}},\ \bibinfo {pages} {42} (\bibinfo {year} {2011})}\BibitemShut
  {NoStop}%
\bibitem [{\citenamefont {Buess}\ \emph {et~al.}(1993)\citenamefont {Buess},
  \citenamefont {Garroway}, \citenamefont {Miller},\ and\ \citenamefont
  {Yesinowski}}]{buess1993explosives}%
  \BibitemOpen
  \bibfield  {author} {\bibinfo {author} {\bibfnamefont {M.}~\bibnamefont
  {Buess}}, \bibinfo {author} {\bibfnamefont {A.}~\bibnamefont {Garroway}},
  \bibinfo {author} {\bibfnamefont {J.}~\bibnamefont {Miller}}, \ and\ \bibinfo
  {author} {\bibfnamefont {J.}~\bibnamefont {Yesinowski}},\ }in\ \href@noop {}
  {\emph {\bibinfo {booktitle} {Advances in Analysis and Detection of
  Explosives}}}\ (\bibinfo  {publisher} {Springer},\ \bibinfo {year} {1993})\
  pp.\ \bibinfo {pages} {361--368}\BibitemShut {NoStop}%
\bibitem [{\citenamefont {Smith}(1995)}]{smith1995nitrogen}%
  \BibitemOpen
  \bibfield  {author} {\bibinfo {author} {\bibfnamefont {J.}~\bibnamefont
  {Smith}},\ }\href {\doibase 10.1049/cp:19950514} {\bibfield  {journal}
  {\bibinfo  {journal} {IET Digital Lib.}\ }\textbf {\bibinfo {volume}
  {European Convention on Security and Detection}},\ \bibinfo {pages} {288 –
  292} (\bibinfo {year} {1995})}\BibitemShut {NoStop}%
\bibitem [{\citenamefont {Mozjoukhine}(2000)}]{mozjoukhine2000two}%
  \BibitemOpen
  \bibfield  {author} {\bibinfo {author} {\bibfnamefont {G.}~\bibnamefont
  {Mozjoukhine}},\ }\href {\doibase 10.1007/BF03162299} {\bibfield  {journal}
  {\bibinfo  {journal} {Appl. Mag. Reson.}\ }\textbf {\bibinfo {volume} {18}},\
  \bibinfo {pages} {527} (\bibinfo {year} {2000})}\BibitemShut {NoStop}%
\bibitem [{\citenamefont {Gregorovi{\v{c}}}\ and\ \citenamefont
  {Apih}(2009)}]{gregorovivc2009tnt}%
  \BibitemOpen
  \bibfield  {author} {\bibinfo {author} {\bibfnamefont {A.}~\bibnamefont
  {Gregorovi{\v{c}}}}\ and\ \bibinfo {author} {\bibfnamefont {T.}~\bibnamefont
  {Apih}},\ }\href {\doibase 10.1016/j.jmr.2009.02.011} {\bibfield  {journal}
  {\bibinfo  {journal} {J. Mag. Res.}\ }\textbf {\bibinfo {volume} {198}},\
  \bibinfo {pages} {215} (\bibinfo {year} {2009})}\BibitemShut {NoStop}%
\bibitem [{\citenamefont {Garroway}\ \emph {et~al.}(1994)\citenamefont
  {Garroway}, \citenamefont {Buess}, \citenamefont {Yesinowski},\ and\
  \citenamefont {Miller}}]{garroway1994narcotics}%
  \BibitemOpen
  \bibfield  {author} {\bibinfo {author} {\bibfnamefont {A.~N.}\ \bibnamefont
  {Garroway}}, \bibinfo {author} {\bibfnamefont {M.~L.}\ \bibnamefont {Buess}},
  \bibinfo {author} {\bibfnamefont {J.~P.}\ \bibnamefont {Yesinowski}}, \ and\
  \bibinfo {author} {\bibfnamefont {J.~B.}\ \bibnamefont {Miller}},\ }in\
  \href@noop {} {\emph {\bibinfo {booktitle} {Substance Detection Systems}}},\
  Vol.\ \bibinfo {volume} {2092}\ (\bibinfo {organization} {International
  Society for Optics and Photonics},\ \bibinfo {year} {1994})\ pp.\ \bibinfo
  {pages} {318--328}\BibitemShut {NoStop}%
\bibitem [{\citenamefont {Magnuson}\ \emph {et~al.}(2001)\citenamefont
  {Magnuson}, \citenamefont {Moeller}, \citenamefont {Shaw},\ and\
  \citenamefont {Sheldon}}]{magnuson2001system}%
  \BibitemOpen
  \bibfield  {author} {\bibinfo {author} {\bibfnamefont {E.~E.}\ \bibnamefont
  {Magnuson}}, \bibinfo {author} {\bibfnamefont {C.~R.}\ \bibnamefont
  {Moeller}}, \bibinfo {author} {\bibfnamefont {J.~D.}\ \bibnamefont {Shaw}}, \
  and\ \bibinfo {author} {\bibfnamefont {A.~G.}\ \bibnamefont {Sheldon}},\
  }\href@noop {} {\enquote {\bibinfo {title} {System and method for contraband
  detection using nuclear quadrupole resonance},}\ } (\bibinfo {year} {2001}),\
  \bibinfo {note} {uS Patent 6,194,898}\BibitemShut {NoStop}%
\bibitem [{\citenamefont {Marino}\ \emph {et~al.}(1982)\citenamefont {Marino},
  \citenamefont {Connors},\ and\ \citenamefont {Leonard}}]{marino1982nitrogen}%
  \BibitemOpen
  \bibfield  {author} {\bibinfo {author} {\bibfnamefont {R.~A.}\ \bibnamefont
  {Marino}}, \bibinfo {author} {\bibfnamefont {R.~F.}\ \bibnamefont {Connors}},
  \ and\ \bibinfo {author} {\bibfnamefont {L.}~\bibnamefont {Leonard}},\
  }\href@noop {} {\emph {\bibinfo {title} {Nitrogen-14 NQR study of energetic
  materials}}},\ \bibinfo {type} {Tech. Rep.}\ (\bibinfo  {institution} {Block
  Engineering, Inc., Cambridge, MA},\ \bibinfo {year} {1982})\BibitemShut
  {NoStop}%
\bibitem [{\citenamefont {Ariando}\ \emph {et~al.}(2019)\citenamefont
  {Ariando}, \citenamefont {Chen}, \citenamefont {Greer},\ and\ \citenamefont
  {Mandal}}]{ariando2019autonomous}%
  \BibitemOpen
  \bibfield  {author} {\bibinfo {author} {\bibfnamefont {D.}~\bibnamefont
  {Ariando}}, \bibinfo {author} {\bibfnamefont {C.}~\bibnamefont {Chen}},
  \bibinfo {author} {\bibfnamefont {M.}~\bibnamefont {Greer}}, \ and\ \bibinfo
  {author} {\bibfnamefont {S.}~\bibnamefont {Mandal}},\ }\href {\doibase
  10.1016/j.jmr.2018.12.007} {\bibfield  {journal} {\bibinfo  {journal} {J.
  Mag. Reson.}\ }\textbf {\bibinfo {volume} {299}},\ \bibinfo {pages} {74}
  (\bibinfo {year} {2019})}\BibitemShut {NoStop}%
\bibitem [{\citenamefont {Vij}(2007)}]{vij2007handbook}%
  \BibitemOpen
  \bibfield  {author} {\bibinfo {author} {\bibfnamefont {D.}~\bibnamefont
  {Vij}},\ }\href@noop {} {\emph {\bibinfo {title} {Handbook of applied solid
  state spectroscopy}}}\ (\bibinfo  {publisher} {Springer Science \& Business
  Media},\ \bibinfo {year} {2007})\BibitemShut {NoStop}%
\bibitem [{\citenamefont {Lucken}(1969)}]{lucken1969nuclear}%
  \BibitemOpen
  \bibfield  {author} {\bibinfo {author} {\bibfnamefont {E.~A.}\ \bibnamefont
  {Lucken}},\ }\href@noop {} {\emph {\bibinfo {title} {Nuclear Quadrupole
  Coupling Constants}}}\ (\bibinfo  {publisher} {Academic Press},\ \bibinfo
  {year} {1969})\BibitemShut {NoStop}%
\bibitem [{\citenamefont {Bonhomme}\ \emph {et~al.}(2010)\citenamefont
  {Bonhomme}, \citenamefont {Gervais}, \citenamefont {Coelho}, \citenamefont
  {Pourpoint}, \citenamefont {Aza{\"\i}s}, \citenamefont {Bonhomme-Coury},
  \citenamefont {Babonneau}, \citenamefont {Jacob}, \citenamefont {Ferrari},
  \citenamefont {Canet} \emph {et~al.}}]{bonhomme2010new}%
  \BibitemOpen
  \bibfield  {author} {\bibinfo {author} {\bibfnamefont {C.}~\bibnamefont
  {Bonhomme}}, \bibinfo {author} {\bibfnamefont {C.}~\bibnamefont {Gervais}},
  \bibinfo {author} {\bibfnamefont {C.}~\bibnamefont {Coelho}}, \bibinfo
  {author} {\bibfnamefont {F.}~\bibnamefont {Pourpoint}}, \bibinfo {author}
  {\bibfnamefont {T.}~\bibnamefont {Aza{\"\i}s}}, \bibinfo {author}
  {\bibfnamefont {L.}~\bibnamefont {Bonhomme-Coury}}, \bibinfo {author}
  {\bibfnamefont {F.}~\bibnamefont {Babonneau}}, \bibinfo {author}
  {\bibfnamefont {G.}~\bibnamefont {Jacob}}, \bibinfo {author} {\bibfnamefont
  {M.}~\bibnamefont {Ferrari}}, \bibinfo {author} {\bibfnamefont
  {D.}~\bibnamefont {Canet}},  \emph {et~al.},\ }\href {\doibase
  10.1002/mrc.2635} {\bibfield  {journal} {\bibinfo  {journal} {Mag. Reson,
  Chem.}\ }\textbf {\bibinfo {volume} {48}},\ \bibinfo {pages} {S86} (\bibinfo
  {year} {2010})}\BibitemShut {NoStop}%
\bibitem [{\citenamefont {Pickard}\ and\ \citenamefont
  {Mauri}(2001)}]{pickard2001all}%
  \BibitemOpen
  \bibfield  {author} {\bibinfo {author} {\bibfnamefont {C.~J.}\ \bibnamefont
  {Pickard}}\ and\ \bibinfo {author} {\bibfnamefont {F.}~\bibnamefont
  {Mauri}},\ }\href {\doibase 10.1103/PhysRevB.63.245101} {\bibfield  {journal}
  {\bibinfo  {journal} {Phys. Rev. B}\ }\textbf {\bibinfo {volume} {63}},\
  \bibinfo {pages} {245101} (\bibinfo {year} {2001})}\BibitemShut {NoStop}%
\bibitem [{\citenamefont {Yates}\ and\ \citenamefont
  {Pickard}(2008)}]{yates2007computations}%
  \BibitemOpen
  \bibfield  {author} {\bibinfo {author} {\bibfnamefont {J.~R.}\ \bibnamefont
  {Yates}}\ and\ \bibinfo {author} {\bibfnamefont {C.~J.}\ \bibnamefont
  {Pickard}},\ }in\ \href {\doibase 10.1002/9780470034590.emrstm1009} {\emph
  {\bibinfo {booktitle} {Encyclopedia of Magnetic Resonance}}},\ \bibinfo
  {editor} {edited by\ \bibinfo {editor} {\bibfnamefont {R.~K.}\ \bibnamefont
  {Harris}}, \bibinfo {editor} {\bibfnamefont {R.}~\bibnamefont {Wasylischen}},
  \ and\ \bibinfo {editor} {\bibfnamefont {M.~J.}\ \bibnamefont {Dauer}}}\
  (\bibinfo  {publisher} {Wiley},\ \bibinfo {address} {New York},\ \bibinfo
  {year} {2008})\BibitemShut {NoStop}%
\bibitem [{\citenamefont {Charpentier}(2011)}]{charpentier2011paw}%
  \BibitemOpen
  \bibfield  {author} {\bibinfo {author} {\bibfnamefont {T.}~\bibnamefont
  {Charpentier}},\ }\href {\doibase 10.1016/j.ssnmr.2011.04.006} {\bibfield
  {journal} {\bibinfo  {journal} {Solid state Nucl. Mag. Reson.}\ }\textbf
  {\bibinfo {volume} {40}},\ \bibinfo {pages} {1} (\bibinfo {year}
  {2011})}\BibitemShut {NoStop}%
\bibitem [{\citenamefont {Bonhomme}\ \emph {et~al.}(2012)\citenamefont
  {Bonhomme}, \citenamefont {Gervais}, \citenamefont {Babonneau}, \citenamefont
  {Coelho}, \citenamefont {Pourpoint}, \citenamefont {Azais}, \citenamefont
  {Ashbrook}, \citenamefont {Griffin}, \citenamefont {Yates}, \citenamefont
  {Mauri} \emph {et~al.}}]{bonhomme2012first}%
  \BibitemOpen
  \bibfield  {author} {\bibinfo {author} {\bibfnamefont {C.}~\bibnamefont
  {Bonhomme}}, \bibinfo {author} {\bibfnamefont {C.}~\bibnamefont {Gervais}},
  \bibinfo {author} {\bibfnamefont {F.}~\bibnamefont {Babonneau}}, \bibinfo
  {author} {\bibfnamefont {C.}~\bibnamefont {Coelho}}, \bibinfo {author}
  {\bibfnamefont {F.}~\bibnamefont {Pourpoint}}, \bibinfo {author}
  {\bibfnamefont {T.}~\bibnamefont {Azais}}, \bibinfo {author} {\bibfnamefont
  {S.~E.}\ \bibnamefont {Ashbrook}}, \bibinfo {author} {\bibfnamefont {J.~M.}\
  \bibnamefont {Griffin}}, \bibinfo {author} {\bibfnamefont {J.~R.}\
  \bibnamefont {Yates}}, \bibinfo {author} {\bibfnamefont {F.}~\bibnamefont
  {Mauri}},  \emph {et~al.},\ }\href {\doibase 10.1021/cr300108a} {\bibfield
  {journal} {\bibinfo  {journal} {Chem. Rev.}\ }\textbf {\bibinfo {volume}
  {112}},\ \bibinfo {pages} {5733} (\bibinfo {year} {2012})}\BibitemShut
  {NoStop}%
\bibitem [{\citenamefont {Socha}\ \emph {et~al.}(2017)\citenamefont {Socha},
  \citenamefont {Hodgkinson}, \citenamefont {Widdifield}, \citenamefont
  {Yates},\ and\ \citenamefont {Dracinsky}}]{socha2017exploring}%
  \BibitemOpen
  \bibfield  {author} {\bibinfo {author} {\bibfnamefont {O.}~\bibnamefont
  {Socha}}, \bibinfo {author} {\bibfnamefont {P.}~\bibnamefont {Hodgkinson}},
  \bibinfo {author} {\bibfnamefont {C.~M.}\ \bibnamefont {Widdifield}},
  \bibinfo {author} {\bibfnamefont {J.~R.}\ \bibnamefont {Yates}}, \ and\
  \bibinfo {author} {\bibfnamefont {M.}~\bibnamefont {Dracinsky}},\ }\href
  {\doibase 10.1021/acs.jpca.7b02810} {\bibfield  {journal} {\bibinfo
  {journal} {J. Phys. Chem. A}\ }\textbf {\bibinfo {volume} {121}},\ \bibinfo
  {pages} {4103} (\bibinfo {year} {2017})}\BibitemShut {NoStop}%
\bibitem [{\citenamefont {Milinkovi{\'c}}\ and\ \citenamefont
  {Bilalbegovi{\'c}}(2012)}]{milinkovic2012nmr}%
  \BibitemOpen
  \bibfield  {author} {\bibinfo {author} {\bibfnamefont {M.}~\bibnamefont
  {Milinkovi{\'c}}}\ and\ \bibinfo {author} {\bibfnamefont {G.}~\bibnamefont
  {Bilalbegovi{\'c}}},\ }\href {\doibase 10.1016/j.cplett.2012.02.043}
  {\bibfield  {journal} {\bibinfo  {journal} {Chem. Phys. Lett.}\ }\textbf
  {\bibinfo {volume} {531}},\ \bibinfo {pages} {105} (\bibinfo {year}
  {2012})}\BibitemShut {NoStop}%
\bibitem [{\citenamefont {Peri{\'c}}\ \emph {et~al.}(2014)\citenamefont
  {Peri{\'c}}, \citenamefont {Gautier}, \citenamefont {Pickard}, \citenamefont
  {Bosio{\v{c}}i{\'c}}, \citenamefont {Grbi{\'c}},\ and\ \citenamefont
  {Po{\v{z}}ek}}]{peric2014solid}%
  \BibitemOpen
  \bibfield  {author} {\bibinfo {author} {\bibfnamefont {B.}~\bibnamefont
  {Peri{\'c}}}, \bibinfo {author} {\bibfnamefont {R.}~\bibnamefont {Gautier}},
  \bibinfo {author} {\bibfnamefont {C.~J.}\ \bibnamefont {Pickard}}, \bibinfo
  {author} {\bibfnamefont {M.}~\bibnamefont {Bosio{\v{c}}i{\'c}}}, \bibinfo
  {author} {\bibfnamefont {M.~S.}\ \bibnamefont {Grbi{\'c}}}, \ and\ \bibinfo
  {author} {\bibfnamefont {M.}~\bibnamefont {Po{\v{z}}ek}},\ }\href {\doibase
  10.1016/j.ssnmr.2014.02.001} {\bibfield  {journal} {\bibinfo  {journal}
  {Solid State Nucl. Mag. Res.}\ }\textbf {\bibinfo {volume} {59}},\ \bibinfo
  {pages} {20} (\bibinfo {year} {2014})}\BibitemShut {NoStop}%
\bibitem [{\citenamefont {Leroy}\ \emph {et~al.}(2019)\citenamefont {Leroy},
  \citenamefont {Johannson},\ and\ \citenamefont {Bryce}}]{leroy2019121}%
  \BibitemOpen
  \bibfield  {author} {\bibinfo {author} {\bibfnamefont {C.}~\bibnamefont
  {Leroy}}, \bibinfo {author} {\bibfnamefont {R.}~\bibnamefont {Johannson}}, \
  and\ \bibinfo {author} {\bibfnamefont {D.~L.}\ \bibnamefont {Bryce}},\ }\href
  {\doibase 10.1021/acs.jpca.8b11490} {\bibfield  {journal} {\bibinfo
  {journal} {J. Phys. Chem. A}\ } (\bibinfo {year} {2019}),\
  10.1021/acs.jpca.8b11490}\BibitemShut {NoStop}%
\bibitem [{\citenamefont {Lee}\ \emph {et~al.}(2007)\citenamefont {Lee},
  \citenamefont {Bing{\"o}l}, \citenamefont {Murakhtina}, \citenamefont
  {Sebastiani}, \citenamefont {Meyer}, \citenamefont {Wegner},\ and\
  \citenamefont {Spiess}}]{lee2007high}%
  \BibitemOpen
  \bibfield  {author} {\bibinfo {author} {\bibfnamefont {Y.~J.}\ \bibnamefont
  {Lee}}, \bibinfo {author} {\bibfnamefont {B.}~\bibnamefont {Bing{\"o}l}},
  \bibinfo {author} {\bibfnamefont {T.}~\bibnamefont {Murakhtina}}, \bibinfo
  {author} {\bibfnamefont {D.}~\bibnamefont {Sebastiani}}, \bibinfo {author}
  {\bibfnamefont {W.~H.}\ \bibnamefont {Meyer}}, \bibinfo {author}
  {\bibfnamefont {G.}~\bibnamefont {Wegner}}, \ and\ \bibinfo {author}
  {\bibfnamefont {H.~W.}\ \bibnamefont {Spiess}},\ }\href {\doibase
  10.1021/jp072112j} {\bibfield  {journal} {\bibinfo  {journal} {J. Phys. Chem.
  B}\ }\textbf {\bibinfo {volume} {111}},\ \bibinfo {pages} {9711} (\bibinfo
  {year} {2007})}\BibitemShut {NoStop}%
\bibitem [{\citenamefont {Rossano}\ \emph {et~al.}(2005)\citenamefont
  {Rossano}, \citenamefont {Mauri}, \citenamefont {Pickard},\ and\
  \citenamefont {Farnan}}]{rossano2005first}%
  \BibitemOpen
  \bibfield  {author} {\bibinfo {author} {\bibfnamefont {S.}~\bibnamefont
  {Rossano}}, \bibinfo {author} {\bibfnamefont {F.}~\bibnamefont {Mauri}},
  \bibinfo {author} {\bibfnamefont {C.~J.}\ \bibnamefont {Pickard}}, \ and\
  \bibinfo {author} {\bibfnamefont {I.}~\bibnamefont {Farnan}},\ }\href
  {\doibase 10.1021/jp044251w} {\bibfield  {journal} {\bibinfo  {journal} {J.
  Phys. Chem. B}\ }\textbf {\bibinfo {volume} {109}},\ \bibinfo {pages} {7245}
  (\bibinfo {year} {2005})}\BibitemShut {NoStop}%
\bibitem [{\citenamefont {Tang}\ and\ \citenamefont
  {Case}(2007)}]{tang2007vibrational}%
  \BibitemOpen
  \bibfield  {author} {\bibinfo {author} {\bibfnamefont {S.}~\bibnamefont
  {Tang}}\ and\ \bibinfo {author} {\bibfnamefont {D.~A.}\ \bibnamefont
  {Case}},\ }\href {\doibase 10.1007/s10858-007-9164-8} {\bibfield  {journal}
  {\bibinfo  {journal} {J Biomolec. NMR}\ }\textbf {\bibinfo {volume} {38}},\
  \bibinfo {pages} {255} (\bibinfo {year} {2007})}\BibitemShut {NoStop}%
\bibitem [{\citenamefont {Bagno}\ \emph {et~al.}(2007)\citenamefont {Bagno},
  \citenamefont {D'Amico},\ and\ \citenamefont {Saielli}}]{bagno2007computing}%
  \BibitemOpen
  \bibfield  {author} {\bibinfo {author} {\bibfnamefont {A.}~\bibnamefont
  {Bagno}}, \bibinfo {author} {\bibfnamefont {F.}~\bibnamefont {D'Amico}}, \
  and\ \bibinfo {author} {\bibfnamefont {G.}~\bibnamefont {Saielli}},\ }\href
  {\doibase 10.1002/cphc.200600725} {\bibfield  {journal} {\bibinfo  {journal}
  {ChemPhysChem}\ }\textbf {\bibinfo {volume} {8}},\ \bibinfo {pages} {873}
  (\bibinfo {year} {2007})}\BibitemShut {NoStop}%
\bibitem [{\citenamefont {Waller}\ \emph {et~al.}(2008)\citenamefont {Waller},
  \citenamefont {Geethalakshmi},\ and\ \citenamefont
  {B{\"u}hl}}]{waller200851v}%
  \BibitemOpen
  \bibfield  {author} {\bibinfo {author} {\bibfnamefont {M.~P.}\ \bibnamefont
  {Waller}}, \bibinfo {author} {\bibfnamefont {K.}~\bibnamefont
  {Geethalakshmi}}, \ and\ \bibinfo {author} {\bibfnamefont {M.}~\bibnamefont
  {B{\"u}hl}},\ }\href {\doibase 10.1021/jp800580n} {\bibfield  {journal}
  {\bibinfo  {journal} {J. Phys. Chem. B}\ }\textbf {\bibinfo {volume} {112}},\
  \bibinfo {pages} {5813} (\bibinfo {year} {2008})}\BibitemShut {NoStop}%
\bibitem [{\citenamefont {R{\"o}hrig}\ and\ \citenamefont
  {Sebastiani}(2008)}]{rohrig2008nmr}%
  \BibitemOpen
  \bibfield  {author} {\bibinfo {author} {\bibfnamefont {U.~F.}\ \bibnamefont
  {R{\"o}hrig}}\ and\ \bibinfo {author} {\bibfnamefont {D.}~\bibnamefont
  {Sebastiani}},\ }\href {\doibase 10.1021/jp075662q} {\bibfield  {journal}
  {\bibinfo  {journal} {J. Phys. Chem. A}\ }\textbf {\bibinfo {volume} {112}},\
  \bibinfo {pages} {1267} (\bibinfo {year} {2008})}\BibitemShut {NoStop}%
\bibitem [{\citenamefont {Kongsted}\ \emph {et~al.}(2007)\citenamefont
  {Kongsted}, \citenamefont {Nielsen}, \citenamefont {Mikkelsen}, \citenamefont
  {Christiansen},\ and\ \citenamefont {Ruud}}]{kongsted2007nuclear}%
  \BibitemOpen
  \bibfield  {author} {\bibinfo {author} {\bibfnamefont {J.}~\bibnamefont
  {Kongsted}}, \bibinfo {author} {\bibfnamefont {C.~B.}\ \bibnamefont
  {Nielsen}}, \bibinfo {author} {\bibfnamefont {K.~V.}\ \bibnamefont
  {Mikkelsen}}, \bibinfo {author} {\bibfnamefont {O.}~\bibnamefont
  {Christiansen}}, \ and\ \bibinfo {author} {\bibfnamefont {K.}~\bibnamefont
  {Ruud}},\ }\href {\doibase 10.1063/1.2424713} {\bibfield  {journal} {\bibinfo
   {journal} {J. Chem. Phys.}\ }\textbf {\bibinfo {volume} {126}},\ \bibinfo
  {pages} {034510} (\bibinfo {year} {2007})}\BibitemShut {NoStop}%
\bibitem [{\citenamefont {Straka}\ \emph {et~al.}(2008)\citenamefont {Straka},
  \citenamefont {Lantto},\ and\ \citenamefont {Vaara}}]{straka2008toward}%
  \BibitemOpen
  \bibfield  {author} {\bibinfo {author} {\bibfnamefont {M.}~\bibnamefont
  {Straka}}, \bibinfo {author} {\bibfnamefont {P.}~\bibnamefont {Lantto}}, \
  and\ \bibinfo {author} {\bibfnamefont {J.}~\bibnamefont {Vaara}},\ }\href
  {\doibase 10.1021/jp711674y} {\bibfield  {journal} {\bibinfo  {journal} {J.
  Phys. Chem. A}\ }\textbf {\bibinfo {volume} {112}},\ \bibinfo {pages} {2658}
  (\bibinfo {year} {2008})}\BibitemShut {NoStop}%
\bibitem [{\citenamefont {Harding}\ \emph {et~al.}(2008)\citenamefont
  {Harding}, \citenamefont {Lenhart}, \citenamefont {Auer},\ and\ \citenamefont
  {Gauss}}]{harding2008quantitative}%
  \BibitemOpen
  \bibfield  {author} {\bibinfo {author} {\bibfnamefont {M.~E.}\ \bibnamefont
  {Harding}}, \bibinfo {author} {\bibfnamefont {M.}~\bibnamefont {Lenhart}},
  \bibinfo {author} {\bibfnamefont {A.~A.}\ \bibnamefont {Auer}}, \ and\
  \bibinfo {author} {\bibfnamefont {J.}~\bibnamefont {Gauss}},\ }\href
  {\doibase 10.1063/1.2943145} {\bibfield  {journal} {\bibinfo  {journal} {J.
  Chem. Phys.}\ }\textbf {\bibinfo {volume} {128}},\ \bibinfo {pages} {244111}
  (\bibinfo {year} {2008})}\BibitemShut {NoStop}%
\bibitem [{\citenamefont {Dumez}\ and\ \citenamefont
  {Pickard}(2009)}]{dumez2009calculation}%
  \BibitemOpen
  \bibfield  {author} {\bibinfo {author} {\bibfnamefont {J.-N.}\ \bibnamefont
  {Dumez}}\ and\ \bibinfo {author} {\bibfnamefont {C.~J.}\ \bibnamefont
  {Pickard}},\ }\href {\doibase 10.1063/1.3081630} {\bibfield  {journal}
  {\bibinfo  {journal} {J. Phys. Chem.}\ }\textbf {\bibinfo {volume} {130}},\
  \bibinfo {pages} {104701} (\bibinfo {year} {2009})}\BibitemShut {NoStop}%
\bibitem [{\citenamefont {Giannozzi}\ \emph {et~al.}(2009)\citenamefont
  {Giannozzi}, \citenamefont {Baroni}, \citenamefont {Bonini}, \citenamefont
  {Calandra}, \citenamefont {Car}, \citenamefont {Cavazzoni}, \citenamefont
  {Ceresoli}, \citenamefont {Chiarotti}, \citenamefont {Cococcioni},
  \citenamefont {Dabo}, \citenamefont {{Dal Corso}}, \citenamefont
  {de~Gironcoli}, \citenamefont {Fabris}, \citenamefont {Fratesi},
  \citenamefont {Gebauer}, \citenamefont {Gerstmann}, \citenamefont
  {Gougoussis}, \citenamefont {Kokalj}, \citenamefont {Lazzeri}, \citenamefont
  {Martin-Samos}, \citenamefont {Marzari}, \citenamefont {Mauri}, \citenamefont
  {Mazzarello}, \citenamefont {Paolini}, \citenamefont {Pasquarello},
  \citenamefont {Paulatto}, \citenamefont {Sbraccia}, \citenamefont {Scandolo},
  \citenamefont {Sclauzero}, \citenamefont {Seitsonen}, \citenamefont
  {Smogunov}, \citenamefont {Umari},\ and\ \citenamefont
  {Wentzcovitch}}]{QE-2009}%
  \BibitemOpen
  \bibfield  {author} {\bibinfo {author} {\bibfnamefont {P.}~\bibnamefont
  {Giannozzi}}, \bibinfo {author} {\bibfnamefont {S.}~\bibnamefont {Baroni}},
  \bibinfo {author} {\bibfnamefont {N.}~\bibnamefont {Bonini}}, \bibinfo
  {author} {\bibfnamefont {M.}~\bibnamefont {Calandra}}, \bibinfo {author}
  {\bibfnamefont {R.}~\bibnamefont {Car}}, \bibinfo {author} {\bibfnamefont
  {C.}~\bibnamefont {Cavazzoni}}, \bibinfo {author} {\bibfnamefont
  {D.}~\bibnamefont {Ceresoli}}, \bibinfo {author} {\bibfnamefont {G.~L.}\
  \bibnamefont {Chiarotti}}, \bibinfo {author} {\bibfnamefont {M.}~\bibnamefont
  {Cococcioni}}, \bibinfo {author} {\bibfnamefont {I.}~\bibnamefont {Dabo}},
  \bibinfo {author} {\bibfnamefont {A.}~\bibnamefont {{Dal Corso}}}, \bibinfo
  {author} {\bibfnamefont {S.}~\bibnamefont {de~Gironcoli}}, \bibinfo {author}
  {\bibfnamefont {S.}~\bibnamefont {Fabris}}, \bibinfo {author} {\bibfnamefont
  {G.}~\bibnamefont {Fratesi}}, \bibinfo {author} {\bibfnamefont
  {R.}~\bibnamefont {Gebauer}}, \bibinfo {author} {\bibfnamefont
  {U.}~\bibnamefont {Gerstmann}}, \bibinfo {author} {\bibfnamefont
  {C.}~\bibnamefont {Gougoussis}}, \bibinfo {author} {\bibfnamefont
  {A.}~\bibnamefont {Kokalj}}, \bibinfo {author} {\bibfnamefont
  {M.}~\bibnamefont {Lazzeri}}, \bibinfo {author} {\bibfnamefont
  {L.}~\bibnamefont {Martin-Samos}}, \bibinfo {author} {\bibfnamefont
  {N.}~\bibnamefont {Marzari}}, \bibinfo {author} {\bibfnamefont
  {F.}~\bibnamefont {Mauri}}, \bibinfo {author} {\bibfnamefont
  {R.}~\bibnamefont {Mazzarello}}, \bibinfo {author} {\bibfnamefont
  {S.}~\bibnamefont {Paolini}}, \bibinfo {author} {\bibfnamefont
  {A.}~\bibnamefont {Pasquarello}}, \bibinfo {author} {\bibfnamefont
  {L.}~\bibnamefont {Paulatto}}, \bibinfo {author} {\bibfnamefont
  {C.}~\bibnamefont {Sbraccia}}, \bibinfo {author} {\bibfnamefont
  {S.}~\bibnamefont {Scandolo}}, \bibinfo {author} {\bibfnamefont
  {G.}~\bibnamefont {Sclauzero}}, \bibinfo {author} {\bibfnamefont {A.~P.}\
  \bibnamefont {Seitsonen}}, \bibinfo {author} {\bibfnamefont {A.}~\bibnamefont
  {Smogunov}}, \bibinfo {author} {\bibfnamefont {P.}~\bibnamefont {Umari}}, \
  and\ \bibinfo {author} {\bibfnamefont {R.~M.}\ \bibnamefont {Wentzcovitch}},\
  }\href {\doibase 10.1088/0953-8984/21/39/395502} {\bibfield  {journal}
  {\bibinfo  {journal} {J. Phys: C.r}\ }\textbf {\bibinfo {volume} {21}},\
  \bibinfo {pages} {395502 (19pp)} (\bibinfo {year} {2009})}\BibitemShut
  {NoStop}%
\bibitem [{\citenamefont {Giannozzi}\ \emph {et~al.}(2017)\citenamefont
  {Giannozzi}, \citenamefont {Andreussi}, \citenamefont {Brumme}, \citenamefont
  {Bunau}, \citenamefont {Nardelli}, \citenamefont {Calandra}, \citenamefont
  {Car}, \citenamefont {Cavazzoni}, \citenamefont {Ceresoli}, \citenamefont
  {Cococcioni}, \citenamefont {Colonna}, \citenamefont {Carnimeo},
  \citenamefont {Corso}, \citenamefont {de~Gironcoli}, \citenamefont {Delugas},
  \citenamefont {Jr}, \citenamefont {Ferretti}, \citenamefont {Floris},
  \citenamefont {Fratesi}, \citenamefont {Fugallo}, \citenamefont {Gebauer},
  \citenamefont {Gerstmann}, \citenamefont {Giustino}, \citenamefont {Gorni},
  \citenamefont {Jia}, \citenamefont {Kawamura}, \citenamefont {Ko},
  \citenamefont {Kokalj}, \citenamefont {K���kbenli}, \citenamefont
  {Lazzeri}, \citenamefont {Marsili}, \citenamefont {Marzari}, \citenamefont
  {Mauri}, \citenamefont {Nguyen}, \citenamefont {Nguyen}, \citenamefont {de-la
  Roza}, \citenamefont {Paulatto}, \citenamefont {Ponc�}, \citenamefont
  {Rocca}, \citenamefont {Sabatini}, \citenamefont {Santra}, \citenamefont
  {Schlipf}, \citenamefont {Seitsonen}, \citenamefont {Smogunov}, \citenamefont
  {Timrov}, \citenamefont {Thonhauser}, \citenamefont {Umari}, \citenamefont
  {Vast}, \citenamefont {Wu},\ and\ \citenamefont {Baroni}}]{QE-2017}%
  \BibitemOpen
  \bibfield  {author} {\bibinfo {author} {\bibfnamefont {P.}~\bibnamefont
  {Giannozzi}}, \bibinfo {author} {\bibfnamefont {O.}~\bibnamefont
  {Andreussi}}, \bibinfo {author} {\bibfnamefont {T.}~\bibnamefont {Brumme}},
  \bibinfo {author} {\bibfnamefont {O.}~\bibnamefont {Bunau}}, \bibinfo
  {author} {\bibfnamefont {M.~B.}\ \bibnamefont {Nardelli}}, \bibinfo {author}
  {\bibfnamefont {M.}~\bibnamefont {Calandra}}, \bibinfo {author}
  {\bibfnamefont {R.}~\bibnamefont {Car}}, \bibinfo {author} {\bibfnamefont
  {C.}~\bibnamefont {Cavazzoni}}, \bibinfo {author} {\bibfnamefont
  {D.}~\bibnamefont {Ceresoli}}, \bibinfo {author} {\bibfnamefont
  {M.}~\bibnamefont {Cococcioni}}, \bibinfo {author} {\bibfnamefont
  {N.}~\bibnamefont {Colonna}}, \bibinfo {author} {\bibfnamefont
  {I.}~\bibnamefont {Carnimeo}}, \bibinfo {author} {\bibfnamefont {A.~D.}\
  \bibnamefont {Corso}}, \bibinfo {author} {\bibfnamefont {S.}~\bibnamefont
  {de~Gironcoli}}, \bibinfo {author} {\bibfnamefont {P.}~\bibnamefont
  {Delugas}}, \bibinfo {author} {\bibfnamefont {R.~A.~D.}\ \bibnamefont {Jr}},
  \bibinfo {author} {\bibfnamefont {A.}~\bibnamefont {Ferretti}}, \bibinfo
  {author} {\bibfnamefont {A.}~\bibnamefont {Floris}}, \bibinfo {author}
  {\bibfnamefont {G.}~\bibnamefont {Fratesi}}, \bibinfo {author} {\bibfnamefont
  {G.}~\bibnamefont {Fugallo}}, \bibinfo {author} {\bibfnamefont
  {R.}~\bibnamefont {Gebauer}}, \bibinfo {author} {\bibfnamefont
  {U.}~\bibnamefont {Gerstmann}}, \bibinfo {author} {\bibfnamefont
  {F.}~\bibnamefont {Giustino}}, \bibinfo {author} {\bibfnamefont
  {T.}~\bibnamefont {Gorni}}, \bibinfo {author} {\bibfnamefont
  {J.}~\bibnamefont {Jia}}, \bibinfo {author} {\bibfnamefont {M.}~\bibnamefont
  {Kawamura}}, \bibinfo {author} {\bibfnamefont {H.-Y.}\ \bibnamefont {Ko}},
  \bibinfo {author} {\bibfnamefont {A.}~\bibnamefont {Kokalj}}, \bibinfo
  {author} {\bibfnamefont {E.}~\bibnamefont {K���kbenli}}, \bibinfo
  {author} {\bibfnamefont {M.}~\bibnamefont {Lazzeri}}, \bibinfo {author}
  {\bibfnamefont {M.}~\bibnamefont {Marsili}}, \bibinfo {author} {\bibfnamefont
  {N.}~\bibnamefont {Marzari}}, \bibinfo {author} {\bibfnamefont
  {F.}~\bibnamefont {Mauri}}, \bibinfo {author} {\bibfnamefont {N.~L.}\
  \bibnamefont {Nguyen}}, \bibinfo {author} {\bibfnamefont {H.-V.}\
  \bibnamefont {Nguyen}}, \bibinfo {author} {\bibfnamefont {A.~O.}\
  \bibnamefont {de-la Roza}}, \bibinfo {author} {\bibfnamefont
  {L.}~\bibnamefont {Paulatto}}, \bibinfo {author} {\bibfnamefont
  {S.}~\bibnamefont {Ponc�}}, \bibinfo {author} {\bibfnamefont
  {D.}~\bibnamefont {Rocca}}, \bibinfo {author} {\bibfnamefont
  {R.}~\bibnamefont {Sabatini}}, \bibinfo {author} {\bibfnamefont
  {B.}~\bibnamefont {Santra}}, \bibinfo {author} {\bibfnamefont
  {M.}~\bibnamefont {Schlipf}}, \bibinfo {author} {\bibfnamefont {A.~P.}\
  \bibnamefont {Seitsonen}}, \bibinfo {author} {\bibfnamefont {A.}~\bibnamefont
  {Smogunov}}, \bibinfo {author} {\bibfnamefont {I.}~\bibnamefont {Timrov}},
  \bibinfo {author} {\bibfnamefont {T.}~\bibnamefont {Thonhauser}}, \bibinfo
  {author} {\bibfnamefont {P.}~\bibnamefont {Umari}}, \bibinfo {author}
  {\bibfnamefont {N.}~\bibnamefont {Vast}}, \bibinfo {author} {\bibfnamefont
  {X.}~\bibnamefont {Wu}}, \ and\ \bibinfo {author} {\bibfnamefont
  {S.}~\bibnamefont {Baroni}},\ }\href {\doibase 10.1088/1361-648X/aa8f79}
  {\bibfield  {journal} {\bibinfo  {journal} {Journal of Physics: Condensed
  Matter}\ }\textbf {\bibinfo {volume} {29}},\ \bibinfo {pages} {465901}
  (\bibinfo {year} {2017})}\BibitemShut {NoStop}%
\bibitem [{\citenamefont {Brown}(1990)}]{brown1990anomalous}%
  \BibitemOpen
  \bibfield  {author} {\bibinfo {author} {\bibfnamefont {R.}~\bibnamefont
  {Brown}},\ }\href {\doibase 10.1515/zna-1990-3-441} {\bibfield  {journal}
  {\bibinfo  {journal} {Zeitschrift f{\"u}r Naturforschung A}\ }\textbf
  {\bibinfo {volume} {45}},\ \bibinfo {pages} {449} (\bibinfo {year}
  {1990})}\BibitemShut {NoStop}%
\bibitem [{\citenamefont {Bayer}(1951)}]{bayer1951theorie}%
  \BibitemOpen
  \bibfield  {author} {\bibinfo {author} {\bibfnamefont {H.}~\bibnamefont
  {Bayer}},\ }\href {\doibase 10.1007/BF01337696} {\bibfield  {journal}
  {\bibinfo  {journal} {Zeit. f{\"u}r Phys.}\ }\textbf {\bibinfo {volume}
  {130}},\ \bibinfo {pages} {227} (\bibinfo {year} {1951})}\BibitemShut
  {NoStop}%
\bibitem [{\citenamefont {Kushida}\ \emph {et~al.}(1956)\citenamefont
  {Kushida}, \citenamefont {Benedek},\ and\ \citenamefont
  {Bloembergen}}]{kushida1956dependence}%
  \BibitemOpen
  \bibfield  {author} {\bibinfo {author} {\bibfnamefont {T.}~\bibnamefont
  {Kushida}}, \bibinfo {author} {\bibfnamefont {G.}~\bibnamefont {Benedek}}, \
  and\ \bibinfo {author} {\bibfnamefont {N.}~\bibnamefont {Bloembergen}},\
  }\href {\doibase 10.1103/PhysRev.104.1364} {\bibfield  {journal} {\bibinfo
  {journal} {Phys. Rev.}\ }\textbf {\bibinfo {volume} {104}},\ \bibinfo {pages}
  {1364} (\bibinfo {year} {1956})}\BibitemShut {NoStop}%
\bibitem [{Note1()}]{Note1}%
  \BibitemOpen
  \bibinfo {note} {In the KBB paper as well as many works, the parameter $C_q$
  is referred to instead by $q = \phi _{ZZ}/e$}\BibitemShut {NoStop}%
\bibitem [{\citenamefont {Brown}(1960)}]{brown1960temperature}%
  \BibitemOpen
  \bibfield  {author} {\bibinfo {author} {\bibfnamefont {R.~J.~C.}\
  \bibnamefont {Brown}},\ }\href {\doibase 10.1063/1.1700882} {\bibfield
  {journal} {\bibinfo  {journal} {J. Phys. Chem.}\ }\textbf {\bibinfo {volume}
  {32}},\ \bibinfo {pages} {116} (\bibinfo {year} {1960})}\BibitemShut
  {NoStop}%
\bibitem [{\citenamefont {Latosi{\'n}ska}\ \emph {et~al.}(2002)\citenamefont
  {Latosi{\'n}ska}, \citenamefont {Kasprzak},\ and\ \citenamefont
  {Utrecht}}]{latosinska2002studies}%
  \BibitemOpen
  \bibfield  {author} {\bibinfo {author} {\bibfnamefont {J.}~\bibnamefont
  {Latosi{\'n}ska}}, \bibinfo {author} {\bibfnamefont {J.}~\bibnamefont
  {Kasprzak}}, \ and\ \bibinfo {author} {\bibfnamefont {R.}~\bibnamefont
  {Utrecht}},\ }\href@noop {} {\bibfield  {journal} {\bibinfo  {journal}
  {Applied Magnetic Resonance}\ }\textbf {\bibinfo {volume} {23}},\ \bibinfo
  {pages} {193} (\bibinfo {year} {2002})}\BibitemShut {NoStop}%
\bibitem [{\citenamefont {Wang}(1955)}]{wang1955pure}%
  \BibitemOpen
  \bibfield  {author} {\bibinfo {author} {\bibfnamefont {T.-C.}\ \bibnamefont
  {Wang}},\ }\href {\doibase 10.1103/PhysRev.99.566} {\bibfield  {journal}
  {\bibinfo  {journal} {Phys. Rev.}\ }\textbf {\bibinfo {volume} {99}},\
  \bibinfo {pages} {566} (\bibinfo {year} {1955})}\BibitemShut {NoStop}%
\bibitem [{\citenamefont {Wheeler}\ and\ \citenamefont
  {Colson}(1976)}]{wheeler1976intermolecular}%
  \BibitemOpen
  \bibfield  {author} {\bibinfo {author} {\bibfnamefont {G.~L.}\ \bibnamefont
  {Wheeler}}\ and\ \bibinfo {author} {\bibfnamefont {S.~D.}\ \bibnamefont
  {Colson}},\ }\href {\doibase 10.1063/1.433231} {\bibfield  {journal}
  {\bibinfo  {journal} {J. Chem. Phys.}\ }\textbf {\bibinfo {volume} {65}},\
  \bibinfo {pages} {1227} (\bibinfo {year} {1976})}\BibitemShut {NoStop}%
\bibitem [{\citenamefont {Estop}\ \emph {et~al.}(1997)\citenamefont {Estop},
  \citenamefont {Alvarez-Larena}, \citenamefont {Belaaraj}, \citenamefont
  {Solans},\ and\ \citenamefont {Labrador}}]{estop1997alpha}%
  \BibitemOpen
  \bibfield  {author} {\bibinfo {author} {\bibfnamefont {E.}~\bibnamefont
  {Estop}}, \bibinfo {author} {\bibfnamefont {A.}~\bibnamefont
  {Alvarez-Larena}}, \bibinfo {author} {\bibfnamefont {A.}~\bibnamefont
  {Belaaraj}}, \bibinfo {author} {\bibfnamefont {X.}~\bibnamefont {Solans}}, \
  and\ \bibinfo {author} {\bibfnamefont {M.}~\bibnamefont {Labrador}},\ }\href
  {\doibase 10.1107/S0108270197007737} {\bibfield  {journal} {\bibinfo
  {journal} {Acta Crystall. C: Crys. Struct. Comm.}\ }\textbf {\bibinfo
  {volume} {53}},\ \bibinfo {pages} {1932} (\bibinfo {year}
  {1997})}\BibitemShut {NoStop}%
\bibitem [{\citenamefont {Deschamps}\ \emph {et~al.}(2011)\citenamefont
  {Deschamps}, \citenamefont {Frisch},\ and\ \citenamefont
  {Parrish}}]{deschamps2011thermal}%
  \BibitemOpen
  \bibfield  {author} {\bibinfo {author} {\bibfnamefont {J.~R.}\ \bibnamefont
  {Deschamps}}, \bibinfo {author} {\bibfnamefont {M.}~\bibnamefont {Frisch}}, \
  and\ \bibinfo {author} {\bibfnamefont {D.}~\bibnamefont {Parrish}},\ }\href
  {\doibase 10.1007/s10870-011-0026-6} {\bibfield  {journal} {\bibinfo
  {journal} {J. Chem.Cryst.}\ }\textbf {\bibinfo {volume} {41}},\ \bibinfo
  {pages} {966} (\bibinfo {year} {2011})}\BibitemShut {NoStop}%
\bibitem [{\citenamefont {Grechishkin}(1959)}]{grechishkin1959nuclear}%
  \BibitemOpen
  \bibfield  {author} {\bibinfo {author} {\bibfnamefont {V.~S.}\ \bibnamefont
  {Grechishkin}},\ }\href@noop {} {\bibfield  {journal} {\bibinfo  {journal}
  {Soviet Physics Uspekhi}\ }\textbf {\bibinfo {volume} {2}},\ \bibinfo {pages}
  {699} (\bibinfo {year} {1959})}\BibitemShut {NoStop}%
\bibitem [{\citenamefont {Vanier}(1960)}]{vanier1960temperature}%
  \BibitemOpen
  \bibfield  {author} {\bibinfo {author} {\bibfnamefont {J.}~\bibnamefont
  {Vanier}},\ }\emph {\bibinfo {title} {Temperature Dependence of Nuclear
  Quadrupole Resonance in KClO$_{3}$}},\ \href@noop {} {Ph.D. thesis},\
  \bibinfo  {school} {Faculty of Graduate studies and Research of McGill
  University} (\bibinfo {year} {1960})\BibitemShut {NoStop}%
\bibitem [{\citenamefont {Zamar}\ and\ \citenamefont
  {Brunetti}(1988)}]{zamar1988uniaxial}%
  \BibitemOpen
  \bibfield  {author} {\bibinfo {author} {\bibfnamefont {R.}~\bibnamefont
  {Zamar}}\ and\ \bibinfo {author} {\bibfnamefont {A.}~\bibnamefont
  {Brunetti}},\ }\href {\doibase 10.1002/pssb.2221500127} {\bibfield  {journal}
  {\bibinfo  {journal} {Phys. Stat. Solidi (B)}\ }\textbf {\bibinfo {volume}
  {150}},\ \bibinfo {pages} {245} (\bibinfo {year} {1988})}\BibitemShut
  {NoStop}%
\bibitem [{\citenamefont {Sullivan}(1971)}]{sullivan1971nuclear}%
  \BibitemOpen
  \bibfield  {author} {\bibinfo {author} {\bibfnamefont {N.}~\bibnamefont
  {Sullivan}},\ }\href {\doibase 10.1063/1.1685130} {\bibfield  {journal}
  {\bibinfo  {journal} {Rev. Sci. Instr.}\ }\textbf {\bibinfo {volume} {42}},\
  \bibinfo {pages} {462} (\bibinfo {year} {1971})}\BibitemShut {NoStop}%
\bibitem [{\citenamefont {Frasson}\ \emph {et~al.}(1959)\citenamefont
  {Frasson}, \citenamefont {Garbuglio},\ and\ \citenamefont
  {Bezzi}}]{frasson1959structure}%
  \BibitemOpen
  \bibfield  {author} {\bibinfo {author} {\bibfnamefont {E.}~\bibnamefont
  {Frasson}}, \bibinfo {author} {\bibfnamefont {C.}~\bibnamefont {Garbuglio}},
  \ and\ \bibinfo {author} {\bibfnamefont {S.}~\bibnamefont {Bezzi}},\ }\href
  {\doibase 10.1107/S0365110X5900038X} {\bibfield  {journal} {\bibinfo
  {journal} {Acta Crystall.}\ }\textbf {\bibinfo {volume} {12}},\ \bibinfo
  {pages} {126} (\bibinfo {year} {1959})}\BibitemShut {NoStop}%
\bibitem [{\citenamefont {Landers}\ \emph {et~al.}(1981)\citenamefont
  {Landers}, \citenamefont {Brill},\ and\ \citenamefont
  {Marino}}]{landers1981HMX}%
  \BibitemOpen
  \bibfield  {author} {\bibinfo {author} {\bibfnamefont {A.}~\bibnamefont
  {Landers}}, \bibinfo {author} {\bibfnamefont {T.}~\bibnamefont {Brill}}, \
  and\ \bibinfo {author} {\bibfnamefont {R.}~\bibnamefont {Marino}},\ }\href
  {\doibase 10.1021/j150618a010} {\bibfield  {journal} {\bibinfo  {journal} {J.
  Phys. Chem.}\ }\textbf {\bibinfo {volume} {85}},\ \bibinfo {pages} {2618}
  (\bibinfo {year} {1981})}\BibitemShut {NoStop}%
\bibitem [{\citenamefont {Buess}\ and\ \citenamefont
  {Caulder}(2004)}]{buess2004factors}%
  \BibitemOpen
  \bibfield  {author} {\bibinfo {author} {\bibfnamefont {M.}~\bibnamefont
  {Buess}}\ and\ \bibinfo {author} {\bibfnamefont {S.}~\bibnamefont
  {Caulder}},\ }\href {\doibase 10.1007/BF03166536} {\bibfield  {journal}
  {\bibinfo  {journal} {Appl. Mag. Reson.}\ }\textbf {\bibinfo {volume} {25}},\
  \bibinfo {pages} {383} (\bibinfo {year} {2004})}\BibitemShut {NoStop}%
\bibitem [{\citenamefont {Allis}\ \emph {et~al.}(2006)\citenamefont {Allis},
  \citenamefont {Prokhorova},\ and\ \citenamefont {Korter}}]{allis2006solid}%
  \BibitemOpen
  \bibfield  {author} {\bibinfo {author} {\bibfnamefont {D.~G.}\ \bibnamefont
  {Allis}}, \bibinfo {author} {\bibfnamefont {D.~A.}\ \bibnamefont
  {Prokhorova}}, \ and\ \bibinfo {author} {\bibfnamefont {T.~M.}\ \bibnamefont
  {Korter}},\ }\href {\doibase 10.1021/jp0554285} {\bibfield  {journal}
  {\bibinfo  {journal} {J. Phys. Chem. A}\ }\textbf {\bibinfo {volume} {110}},\
  \bibinfo {pages} {1951} (\bibinfo {year} {2006})}\BibitemShut {NoStop}%
\bibitem [{\citenamefont {Sewell}\ \emph {et~al.}(2003)\citenamefont {Sewell},
  \citenamefont {Menikoff}, \citenamefont {Bedrov},\ and\ \citenamefont
  {Smith}}]{sewell2003molecular}%
  \BibitemOpen
  \bibfield  {author} {\bibinfo {author} {\bibfnamefont {T.~D.}\ \bibnamefont
  {Sewell}}, \bibinfo {author} {\bibfnamefont {R.}~\bibnamefont {Menikoff}},
  \bibinfo {author} {\bibfnamefont {D.}~\bibnamefont {Bedrov}}, \ and\ \bibinfo
  {author} {\bibfnamefont {G.~D.}\ \bibnamefont {Smith}},\ }\href {\doibase
  10.1063/1.1599273} {\bibfield  {journal} {\bibinfo  {journal} {J. Chem.
  Phys.}\ }\textbf {\bibinfo {volume} {119}},\ \bibinfo {pages} {7417}
  (\bibinfo {year} {2003})}\BibitemShut {NoStop}%
\bibitem [{\citenamefont {Wu}\ \emph {et~al.}(2011)\citenamefont {Wu},
  \citenamefont {Kalia}, \citenamefont {Nakano},\ and\ \citenamefont
  {Vashishta}}]{wu2011vibrational}%
  \BibitemOpen
  \bibfield  {author} {\bibinfo {author} {\bibfnamefont {Z.}~\bibnamefont
  {Wu}}, \bibinfo {author} {\bibfnamefont {R.~K.}\ \bibnamefont {Kalia}},
  \bibinfo {author} {\bibfnamefont {A.}~\bibnamefont {Nakano}}, \ and\ \bibinfo
  {author} {\bibfnamefont {P.}~\bibnamefont {Vashishta}},\ }\href {\doibase
  10.1063/1.3587135} {\bibfield  {journal} {\bibinfo  {journal} {J. Chem.
  Phys.}\ }\textbf {\bibinfo {volume} {134}},\ \bibinfo {pages} {204509}
  (\bibinfo {year} {2011})}\BibitemShut {NoStop}%
\bibitem [{\citenamefont {Groom}\ \emph {et~al.}(2016)\citenamefont {Groom},
  \citenamefont {Bruno}, \citenamefont {Lightfoot},\ and\ \citenamefont
  {Ward}}]{groom2016cambridge}%
  \BibitemOpen
  \bibfield  {author} {\bibinfo {author} {\bibfnamefont {C.~R.}\ \bibnamefont
  {Groom}}, \bibinfo {author} {\bibfnamefont {I.~J.}\ \bibnamefont {Bruno}},
  \bibinfo {author} {\bibfnamefont {M.~P.}\ \bibnamefont {Lightfoot}}, \ and\
  \bibinfo {author} {\bibfnamefont {S.~C.}\ \bibnamefont {Ward}},\ }\href
  {\doibase 10.1107/S2052520616003954} {\bibfield  {journal} {\bibinfo
  {journal} {Acta Crystallographica Section B: Structural Science, Crystal
  Engineering and Materials}\ }\textbf {\bibinfo {volume} {72}},\ \bibinfo
  {pages} {171} (\bibinfo {year} {2016})}\BibitemShut {NoStop}%
\bibitem [{Note2()}]{Note2}%
  \BibitemOpen
  \bibinfo {note} {We disregard the additional line one expects due to the
  heavy isotope $^{37}$Cl, which we did not account for here despite its
  natural abundance of 24.4\%. The NQR frequencies of the two isotopes should
  have the ratio of their respective quadrupole moments, $\nu _{^{35}\protect
  \text {Cl}}/\nu _{^{37}\protect \text {Cl}} = Q_{^{35}\protect \text
  {Cl}}/Q_{^{37}\protect \text {Cl}} = 1.2704$, under normal conditions where
  the nuclei are not subject to significant deformations.}\BibitemShut {Stop}%
\bibitem [{\citenamefont {Moross}\ and\ \citenamefont
  {Story}(1966)}]{moross1966temperature}%
  \BibitemOpen
  \bibfield  {author} {\bibinfo {author} {\bibfnamefont {G.~G.}\ \bibnamefont
  {Moross}}\ and\ \bibinfo {author} {\bibfnamefont {H.~S.}\ \bibnamefont
  {Story}},\ }\href {\doibase 10.1063/1.1728116} {\bibfield  {journal}
  {\bibinfo  {journal} {J. Chem. Phys.}\ }\textbf {\bibinfo {volume} {45}},\
  \bibinfo {pages} {3370} (\bibinfo {year} {1966})}\BibitemShut {NoStop}%
\bibitem [{Note3()}]{Note3}%
  \BibitemOpen
  \bibinfo {note} {NQR measurements involve direct observation of the coupling
  constants $eQ\phi _{ZZ}/h$, whereas in the present study, we leave $Q$ a
  constant and account for dynamical effects on $\phi _{ZZ}$. Values of $Q$ in
  standard tables are of the intrinsic quadrupole moments. But the nuclear
  environment may cause deformations of the nucleus that induce contributions
  to the quadrupole moment.}\BibitemShut {Stop}%
\end{thebibliography}%

\end{document}